\documentstyle[aps,epsf,pra,multicol,amssymb,float]{revtex}

\newcommand{\ket}[1]{\vert#1\rangle}
\newcommand{\bra}[1]{\langle#1\vert}

\newcommand{\projector}[1]{\vert#1\rangle\langle#1\vert}

\title{The physics of forgetting: Landauer's erasure principle and information theory}
\author{M. B. Plenio and V. Vitelli\\
\it Optics Section, The Blackett Laboratory, Imperial College,\\
London SW7 2BW, UK}

\begin{document}
\maketitle

\begin{abstract}
This article discusses the concept of information and its intimate relationship
with physics. After an introduction of all the necessary quantum mechanical and
information theoretical concepts we analyze Landauer's principle that states
that the erasure of information is inevitably accompanied by the generation of
heat. We employ this principle to rederive a number of results in classical and
quantum information theory whose rigorous mathematical derivations are
difficult. This demonstrates the usefulness of Landauer's principle and
provides an introduction to the physical theory of information.
\end{abstract}


\begin{multicols}{2}

\section{Introduction}
\label{introduction}

In recent years great interest in quantum information theory has been generated
by the prospect of employing its laws to design devices of surprising power
\cite{Ekert J 96,Vedral P 98,Steane 98,Hughes ADLMS 95,PhysicsWorld
98,TMRQIT,Plenio V 98}. Ideas include quantum computation \cite{Vedral P
98,PhysicsWorld 98,Shor 99}, quantum teleportation \cite{Plenio V 98,Bennett
BCJPW 93} and quantum cryptography \cite{Hughes ADLMS 95,PhysicsWorld
98,Bennett B 84,Ekert 91}. In this article, we will not deal with such
applications directly, but rather with some of the underlying concepts and
physical principles. Rather than presenting very abstract mathematical proofs
originating from the mathematical theory of information, we will base our
arguments as far as possible on the paradigm that information is physical. In
particular, we are going to employ the fact that the erasure of one bit of
information always increases the thermodynamical entropy of the world by $k
ln2$. This principle, originally suggested by Rolf Landauer in 1961
\cite{Landauer 61,Leff R 90}, has been applied successfully by Charles Bennett
to resolve the notorious Maxwell's demon paradox \cite{Leff R 90,Bennett 82}.
In this article we will argue that Landauer's principle provides a bridge
between information theory and physics and that, as such, it sheds light on a
number of issues regarding classical and quantum information processing and the
truly quantum mechanical feature of entanglement and non-local correlations
\cite{Plenio V 98}. We introduce the basic concepts both at an informal level
as well as a more mathematical level to allow a more thorough understanding of
these concepts. This enables us to approach and answer a number of questions at
the interface between pure physics and technology such as:

\begin{enumerate}

\item What is the greatest amount of classical information we can send
reliably through a noisy classical or quantum channel?

\item Can quantum information be copied and compressed as we do with classical
information on a daily basis?

\item If entanglement is such a useful resource, how much of it can
be extracted from an arbitrary quantum system composed of two
parts by acting locally on each of the two?

\noindent
\end{enumerate}

\noindent

\begin{figure}[htb]
    \centerline{\epsfxsize=15.cm \epsffile{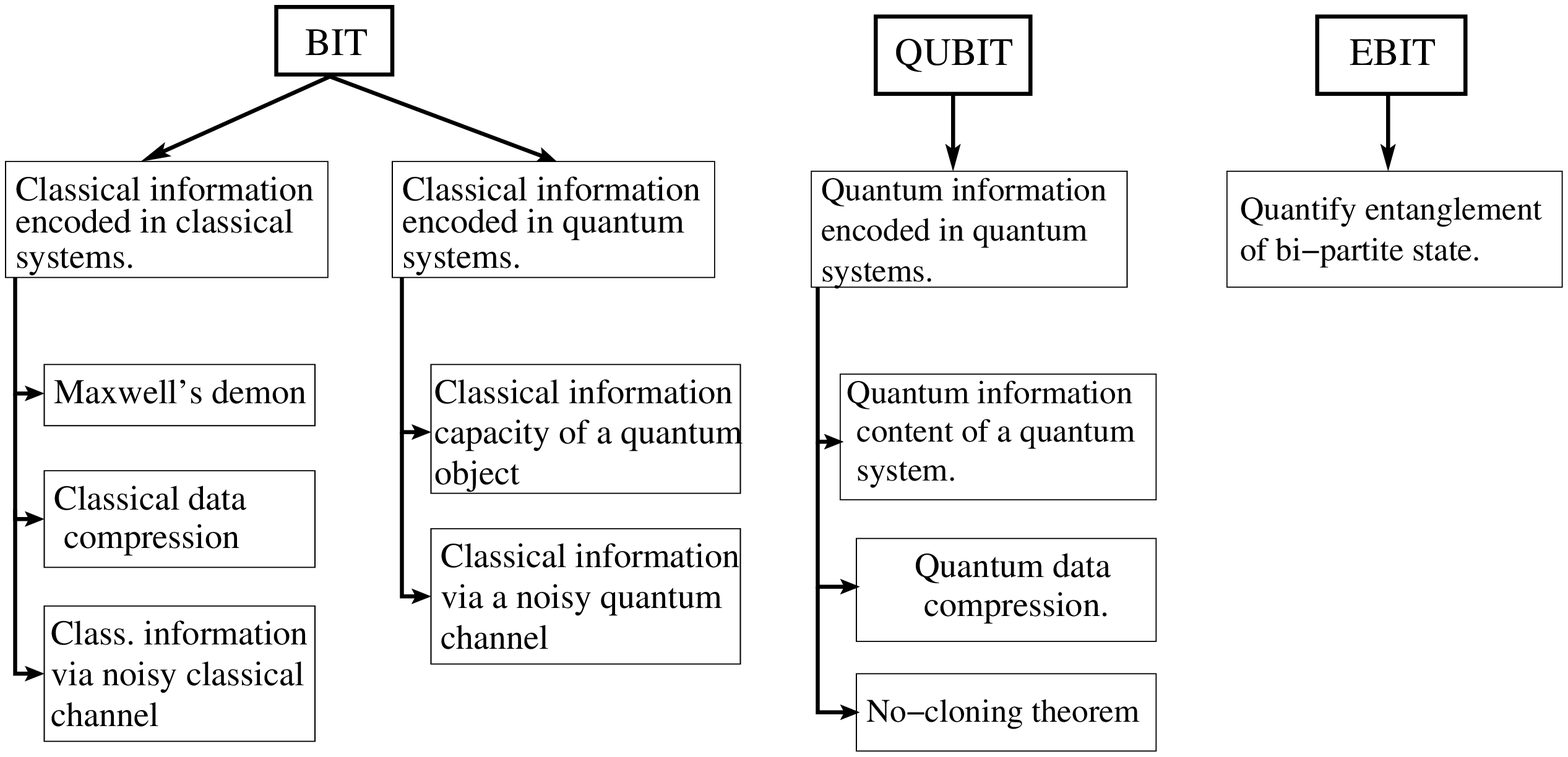}}
    \vspace*{0.15cm}
    \caption{The essential structure of the article is captured in this
    diagram.}
    \label{structure}
\end{figure}

The full meaning of these questions and their answer will gradually emerge
after explaining some of the unpleasant but unavoidable jargon used to state
them. For the time being, our only remark is that Landauer's principle will be
our companion in this journey. A glance at what lies ahead can be readily
obtained by inspecting the "map" of this paper in Fig. \ref{structure}.

\end{multicols}

\begin{multicols}{2}

A final word on the level of this article: the concepts of
entanglement and quantum information are of great importance in
contemporary research on quantum mechanics, but they seldom appear
in graduate textbooks on quantum mechanics. This article, while
making little claim to originality in the sense that it does not
derive new results, tries to fill this gap. It provides an
introduction to the physical theory of information and the concept
of entanglement and is written from the perspective of an advanced
undergraduate student in physics, who is eager to learn, but may
not have the necessary mathematical background to directly access
the original sources. This pedagogical outlook is also reflected
in the choice of particularly readable references mainly textbooks
and lecture notes, that we hope the reader will consult for a more
comprehensive treatment of the advanced topics \cite{Preskill
notes,Plenio notes,Mabuchi notes,Feynmanbook,Peres 95,Nielsen C
00,Cover T 91}. We also try our best to use mathematics as a {\em
language} rather than as a {\em weapon}. Every idea is first
motivated, then illustrated with a non-trivial example and
occasionally extended to the general case by using Landauers
principle. The reader will not be drowned in a sea of indices or
obscure symbols, but he will (hopefully) be guided to work out the
simple examples in parallel with the text. Most of the subtle
concepts in quantum mechanics can indeed be illustrated using
simple matrix manipulations. On the other hand, the choice to
actively involve the reader in calculations makes this article
unsuitable for bed-time readings. In fact, it is a good idea to
keep a pen and plenty of blank paper within reach, while you read
on.

\section{Classical information encoded in classical systems}
\label{Classicalinformationencodedinclassicalsystems}

\subsection{The bit}
\label{Thebit}

In this section we will try to build an intuitive understanding of the concept
of classical information. A more quantitative approach will be taken in section
\ref{Theinformationcontentofaclassicalstateinbits}, but for the full blown
mathematical apparatus we refer the reader to textbooks, e.g. \cite{Cover T
91}.

\begin{figure}[b]
    \centerline{\epsfxsize=10.cm \epsffile{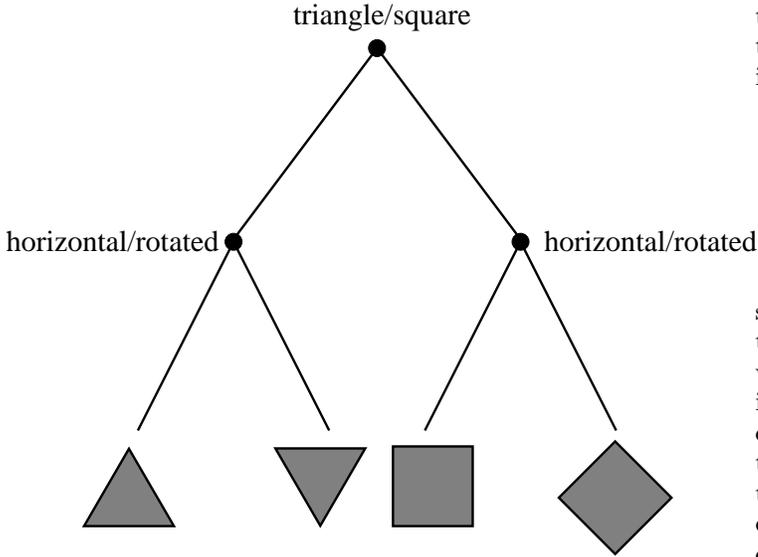}}
    \vspace*{0.15cm}
    \caption{An example for a decision tree. Two binary choices have to be
    made to identify the shape (triangle or square) and the orientation (horizontal or
    rotated). In sending with equal probability one of the four objects, one
    therefore transmits 2 bits of information.}
    \label{decision}
\end{figure}

Imagine that you are holding an object, be it an array of cards,
geometric shapes or a complex molecule and we ask the following
question: {\it what is the information content of this object?} To
answer this question, we introduce another party, say a friend,
who shares some background knowledge with us (e.g. the same
language or other sets of prior agreements that make communication
possible at all), but who does not know the state of the object.
We define the {\it information content} of the object as the size
of the set of instructions that our friend requires to be able to
reconstruct the object, or better the state of the object. For
example, assume that the object is a spin-up particle and that we
share with the friend the background knowledge that the spin is
oriented either upwards or downwards along the {\it z} direction
with equal probability (see fig. \ref{decision} for a slightly
more involved example). In this case, the only instruction we need
to transmit to another party to let him recreate the state is
whether the state is spin-up $\uparrow$ or spin-down $\downarrow$.
This example shows that in some cases the instruction transmitted
to our friend is just a choice between two alternatives. More
generally, we can reduce a complicated set of instructions to $n$
binary choices. If that is done we readily get a measure of the
information content of the object by simply counting the number of
binary choices. In classical information theory, a variable which
can assume only the values $0$ or $1$ is called a {\it bit}.
Instructions to make a binary choice can be given by transmitting
$1$ to suggest one of the alternative (say arrow up $\uparrow$)
and $0$ for the other (arrow down $\downarrow$). To sum up, we say
that $n$ bits of information can be encoded in a system when
instructions in the form of $n$ binary choices need to be
transmitted to identify or recreate the state of the system.

\subsection{Information is physical}
\label{Informationisphysical}

In the previous subsection we have introduced the concept of the bit as the
unit of information. In the course of the argument we mentioned already that
information can be encoded in physical systems. In fact, looking at it more
closely, we realize that any information is encoded, processed and transmitted
by physical means. Physical systems such as capacitors or spins are used for
storage, sound waves or optical fibers for transmission and the laws of
classical mechanics, electrodynamics or quantum mechanics dictate the
properties of these devices and limit our capabilities for information
processing. These rather obvious looking statements, however, have significant
implications for our understanding of the concept of information as they
emphasize that the theory of information is not a purely mathematical concept,
but that the properties of its basic units are dictated by the laws of physics.
The different laws that rule in the classical world and the quantum world for
example results in different information processing capabilities and it is this
insight that sparked the interest in the general field of quantum information
theory.

In the following we would like to further corroborate the view that information
and physics should be unified to a physical theory of information by showing
that the process of erasure of information is invariably accompanied by the
generation of heat and that this insight leads to a resolution of the
longstanding Maxwell demon paradox which is really a prime example of the deep
connection between physics and information. The rest of the article will then
attempt to apply the connection between erasure of information and physical
heat generation further to gain insight into recent results in quantum
information theory.

\subsection{Erasing classical information from classical systems: Landauer's principle}
\label{Erasingclassicalinformationfromclassicalsystems}

\begin{figure}[!htb]
    \centerline{\epsfxsize=10.cm \epsffile{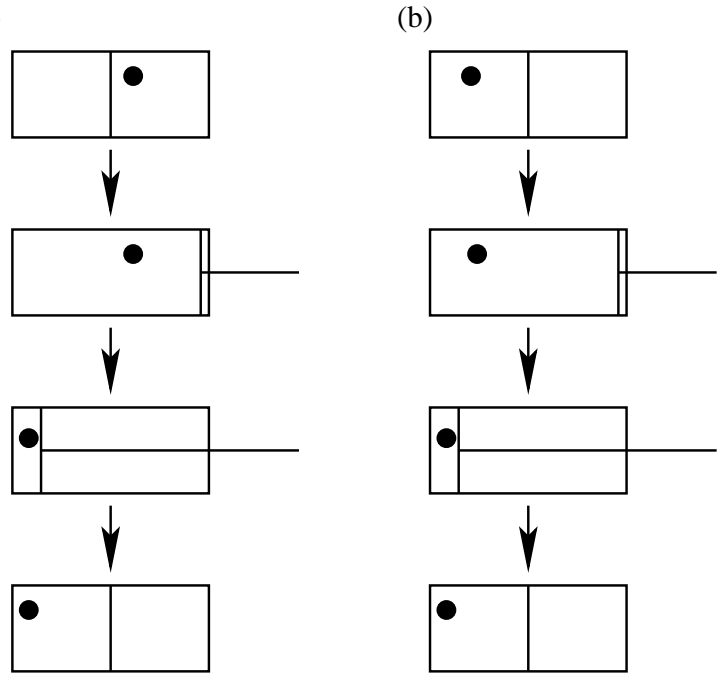}}
    \vspace*{0.15cm}
    \caption{We erase the information of the position of the atom. First we extract
    the wall separating the two halves of the box. Then we use a piston to shift the
    atom to the left side of the box. After the procedure, the atom is on the left
    hand side of the box irrespective of its intial state. Note that the procedure
    has to work irrespective of whether the atom is initially on the right (figure (a))
    or on the left side (figure (b)).  }
    \label{erasure}
\end{figure}

We begin our investigations by concentrating on classical
information. In 1961, Rolf Landauer had the important insight that
there is a fundamental asymmetry in the way Nature allows us to
process information \cite{Landauer 61}. Copying classical
information can be done reversibly and without wasting any energy,
but when information is erased there is always an energy cost of
$kTln2$ per classical bit to be paid. For example, as shown in
fig. \ref{erasure}, we can encode one bit of information in a
binary device composed of a box with a partition.

The box is
filled with a one molecule gas that can be on either side of the
partition, {\it but we do not know which one}. We assume that we
erase the bit of information encoded in the position of the
molecule by extracting the partition and compressing the molecule
in the right part of the box irrespective of where it was before.
We say that information has been erased during the compression
because we will never find out where the molecule was originally.
Any binary message encoded is lost! The physical result of the
compression is a decrease in the thermodynamical entropy of the
gas by $kln2$. The minimum work that we need to do on the box is
$kTln2$, if the compression is isothermal and quasi-static.
Furthermore an amount of heat equal to $kTln2$ is dumped in the
environment at the end of the process.

Landauer's conjectured that this energy/entropy cost cannot be reduced below
this limit irrespective of how the information is encoded and subsequently
erased - it is a fundamental limit. In the discussion of the Maxwell demon in
the next section we will see that this principle can be deduced from the second
law of thermodynamics and is in fact equivalent to it \cite{Schumacher 00}.
Landauer's discovery is important both theoretically and practically as on the
one hand it relates the concept of information to physical quantities like
thermodynamical entropy and free energy and on the other hand it may force the
future designers of quantum devices to take into account the heat production
caused by the erasure of information although this effect is tiny and
negligible in today's technology.

At this point we are ready to summarize our findings on the
physics of classical information.

\begin{center}
\begin{tabular}{|l|}
    \hline
    1) Information is always encoded in a physical system.\\
    2) The erasure of information causes a generation of $kTln2$\\
     of heat per bit in the environment.\\ \hline
\end{tabular}
\end{center}

Armed with this knowledge we will present the first successful
application of the erasure principle: the solution of the
Maxwell's demon paradox that has plagued the foundations of
thermodynamics for almost a century.

\subsection{Maxwell's demon deposed}
\label{Maxwelldemon}

\subsubsection{The paradox}
\label{Theparadox}

\begin{figure}[!htb]
    \centerline{\epsfxsize=9.cm \epsffile{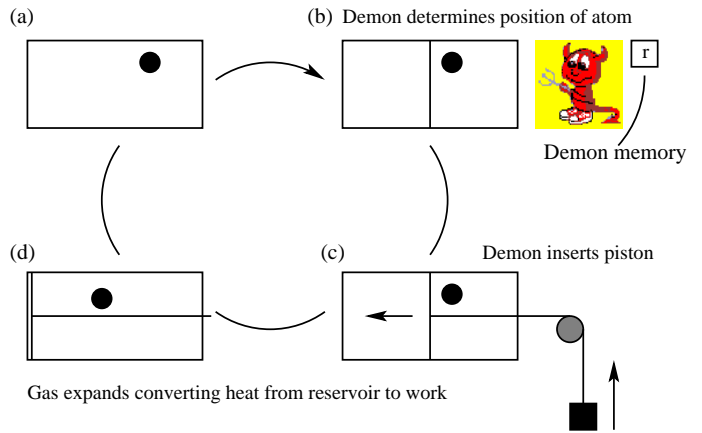}}
    \vspace*{0.15cm}
    \caption{A schematical picture of Szilard's engine of a box filled with a
    one atom gas. Initially the position of the atom is unknown. Then the demon
    measures the position and depending on the outcome inserts a
    piston. Then the gas expands and thereby does work on a load
    attached to the piston. This procedure is repeated and we
    apparently do work at the sole expense of extracting heat from one
    reservoir only.}
    \label{szilard}
\end{figure}

In this section we present a simplified version of the Maxwell's
demon paradox suggested by Leo Szilard in 1929 \cite{Szilard 29}.
It employs an intelligent being or a computer of microscopic size,
operating a heat engine with a single molecule working fluid
(figure \ref{szilard}).

In this scheme, the molecule is originally
placed in a box, free to move in the entire volume V as shown in
step (a). Step (b) consists of inserting a partition which divides
the box in two equal parts. At this point the Maxwell's demon
measures in which side of the box the molecule is and records the
result (in the figure the molecule is pictured on the right-hand
side of the partition as an example). In step (c) the Maxwell
demon uses the information to replace the partition with a piston
and couple the latter to a load. In step (d) the one-molecule gas
is put in contact with a reservoir and expands isothermically to
the original volume $V$. During the expansion the gas draws heat
from the reservoir and does work to lift the load. Apparently the
device is returned to its initial state and it is ready to perform
another cycle whose net result is again full conversion of heat
into work, a process forbidden by the second law of
thermodynamics.

Despite its deceptive simplicity, the argument above has missed an
important point: while the gas in the box has returned to its
initial state, the mind of the demon hasn't! In fact, the demon
needs to erase the information stored in his mind for the process
to be truly cyclic. This is because the information in the brain
of the demon is stored in physical objects and cannot be regarded
as a purely mathematical concept! The first attempts to solve the
paradox had missed this point completely and relied on the
assumption that the act of acquisition of information by the demon
entails an energy cost equal to the work extracted by the demonic
engine, thus preventing the second law to be defeated. This
assumption is wrong! Information on the position of the particle
can be acquired reversibly without having to pay the energy bill,
but erasing information does have a cost! This important remark
was first made by Bennett in a very readable paper on the physics
of computation \cite{Bennett 82}. We will analyze his argument in
some detail. Bennett developed Szilard's earlier suggestion
\cite{Szilard 29} that the demon's mind could be viewed as a
two-state system that stores one bit of information about the
position of the particle. In this sense, the demon's mind can be
an inanimate binary system which represents a significant step
forward, as it rids the discussion from the dubious concept of
intelligence. After the particle in the box is returned to the
initial state the bit of information is still stored in the
demon's mind (ie in the binary device). Consequently, this bit of
information needs to be erased to return the demon's mind to its
initial state. By Landauer's principle this erasure has an energy
cost

\begin{equation}
    W_{erasure} = -kT ln2 \; .
\label{er}
\end{equation}
On the other hand, the work extracted by the demonic engine in the
isothermal expansion is
\begin{equation}
    W_{extracted} = +kT ln2 \; .
\label{extract}
\end{equation}
All the work gained by the engine is needed to erase the information in the
demon's mind, so that no net work is produced in the cycle. Furthermore, the
erasure transfers into the reservoir the same amount of heat that was drawn
from it originally. So there is no net flow of heat either. There is no net
result after the process is completed and the second law of thermodynamics is
saved! The crucial point in Bennett's argument is that the information
processed by the demon must be encoded in a physical system that obeys the laws
of physics. The second law of thermodynamics states that there is no entropy
decrease in a closed system that undergoes a cyclic transformation. Therefore
if we let the demon measure the Szilard's engine we need to include the
physical state he uses to store the information in the analysis, otherwise
there would be an interaction with the environment and the system would not be
closed. One could also view the demon's mind as a heat bath initially at zero
temperature.  After storing information in it, the mind appears to an outside
observer like a random sequence of digits and one could therefore say that the
demons mind has been heated up. Having realized that the demon's mind is a
second heat bath, we now have a perfectly acceptable process that does not
violate the second law of thermodynamics.

\subsubsection{Generalized entropy}
\label{Generalizedentropy}

\begin{figure}[!htb]
    \centerline{\epsfxsize=9.cm \epsffile{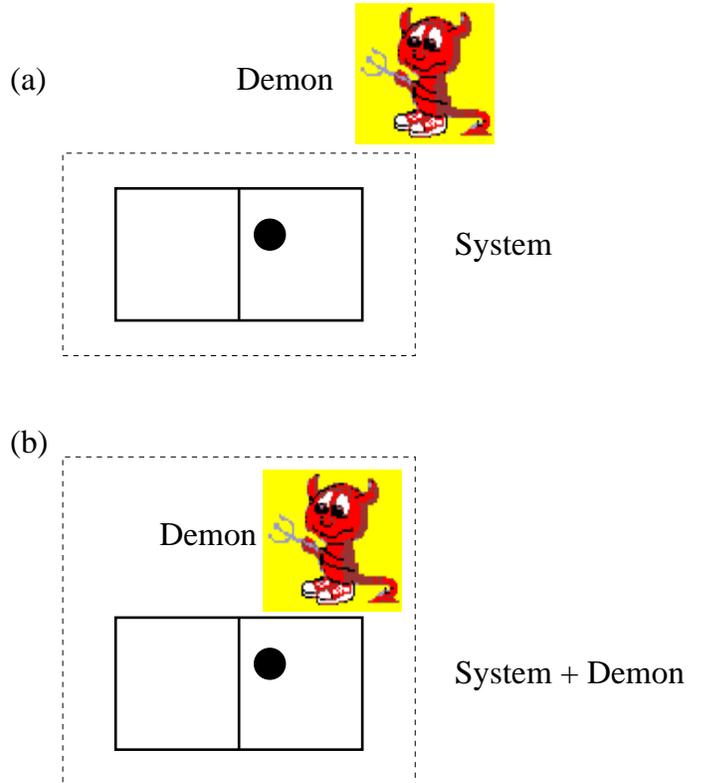}}
    \vspace*{0.15cm}
    \caption{A figure that shows the two different viewpoints discussed in this section.
    The demon is outside the system which consists of the box and the atom only (figure (a))
    or the demon and the box form a joint system that is closed.}
    \label{insideoutside}
\end{figure}

The solution of the paradox presented in the last section views
the "brain of the demon" as a physical system to be included in
the entropy balance together with the box that is being observed
(see part (b) of figure \ref{insideoutside}).

A different approach
can be taken if one does not want to consider explicitly the
workings of the demon's mind, but just treat it as an external
observer that obtains information about the system (see part (a)
of figure \ref{insideoutside}). This is done by including in the
definition of the entropy of the system a term that represents the
knowledge that the demon has on the state of the system together
with the well known term representing how ordered the state is
\cite{Zurek 89,Leff R 90}.

In the context of Szilard's engine we found that the demon extracts from the
engine an amount of work given by
\begin{equation}
    W_{extracted} = kT ln2 = \Delta Q = T\Delta S,
    \label{work ex}
\end{equation}
where $\Delta S$ is the change of thermodynamical entropy in the
system when the heat $\Delta Q$ is absorbed from the environment.
On the other hand, to erase his memory he uses at least an equal
amount of work given by
\begin{equation}
    W_{erasure} = -kT ln2 = -TI,
    \label{information}
\end{equation}
where $I$ denotes the information required by the demon to specify
on which side of the box the molecule is times the scaling factor
$k ln2$. In this case the information is just 1 bit. The scaling
factor is introduced for consistency because the definition of
information is given in bits as a logarithm in basis 2 of the
number of memory levels in the demon's mind.

The total work gained (equal to the total heat exchanged $Q_{total}$ since the
system is kept at constant temperature T) is thus given by
\begin{equation}
    W_{total} = W_{erasure} + W_{extracted} = Q_{total} = T(\Delta S - I) =
    0\; .
    \label{work tot}
\end{equation}
This suggests that the second law of thermodynamics is not
violated if we introduce a generalized definition of entropy $\Im$
(in bits) as the difference of the thermodynamical entropy of the
system $\Delta S$ and the information about the system $I$
possessed by an external observer.

\begin{equation}
    \Im=\Delta S - I \; .
    \label{total entropy}
\end{equation}

The idea of modifying the definition of thermodynamical entropy
that represents an objective property of the physical system with
an "informational term" relative to an external observer appears
bizarre at first sight. Physical properties like entropy identify
and distinguish physical states. By introducing a notion as
information directly in the second law of thermodynamics we
somehow bolster the view that an ensemble composed of partitioned
boxes each containing a molecule in a position unknown to us is
not the same physical state than an ensemble in which we know
exactly on which side of the partition the molecule is in each
box. Why? Because we can extract work from the second state by
virtue of the knowledge we gained, but we cannot do the same with
the first. We will encounter similar arguments in later sections
when we study the concept of information in the context of quantum
theory. For the time being, we remark that the approach presented
in this section to the solution of the Maxwell's demon paradox
adds new meaning to the slogan {\it information is physical}.
Information is physical because it is always encoded in a physical
system and also because the information we possess about a
physical system contributes to define the state of the system.

\subsection{The information content of a classical state in bits}
\label{Theinformationcontentofaclassicalstateinbits}

So far we have discussed how information is encoded in a classical system and
subsequently erased from it. However, we really haven't quantified the
information content of a complicated classical system composed of many
components each of which can be in one of $n$ states with probability $p_n$.
This problem is equivalent to determining the information content of a long
classical message. In fact, a classical message is encoded in a string of
classical objects each representing a letter from a known alphabet occurring
with a certain probability. The agreed relation between objects and letters
represents the required background knowledge for communication. Bob sends this
string of objects to Alice. She knows how the letters of the alphabet are
encoded in the objects, but she does not know the message that Bob is sending.
When Alice receives the objects, she can decode the information in the message,
provided that none of the objects has been accidentally changed on the way to
her. Can we quantify the information transmitted if we know that each letter
$\rho_{i}$ occurs in the message with probability $p_{i}$? Let us begin with
some hand-waving which is followed in the next section by a formally correct
argument. Assume that our alphabet is composed of only two letters $1$ and $0$
occurring with probability $p_1 = 0.1$ and $p_0 = 0.9$ respectively. Suppose we
send a very long message, what is the average information sent per letter?
Naively, one could say that if each letter can be either $1$ or $0$ then the
information transmitted per letter has to be {\it $1$ bit}. But this answer
does not take into account the different probabilities associated with
receiving a $1$ or a $0$. For example, presented with an object Alice can guess
its identity in $90\%$ of the cases by simply assuming it is $0$.  On the other
hand, if the letters $1$ and $0$ come out with equal probability, she will
guess correctly only $50\%$ of the time. Therefore her surprise will usually be
bigger in the second case as she doesn't know what to expect. Let us quantify
Alice's surprise when she finds letter $i$ which normally occurs with
probability $p_i$ by

\begin{equation}
    \mbox{surprise letter i} = log \frac{1}{p_i} \; .
\end{equation}
We have chosen the logarithm of $\frac{1}{p_i}$ because if we guess two
letters, then the surprise should be additive, i.e.
\begin{eqnarray}
    \log(\frac{1}{p_i}\frac{1}{p_j}) &=& \log \frac{1}{p_i} + \log \frac{1}{p_j}\;\;.
    \nonumber\\
    &=& \mbox{surprise letter i} + \mbox{surprise letter j}\;\;.
\end{eqnarray}
and this can only be satisfied by the logarithm. Now we can
compute the average surprise, which we find to be given by the
Shannon entropy

\begin{equation}
    H = \sum_i p_i \log \frac{1}{p_i} = - \sum_i p_i \log p_i \; .
\end{equation}
This argument is of course hand-waving and therefore the next
section addresses the problem more formally by asking how much one
can compress a message, i.e. how much redundancy is included in a
message.

\subsubsection{Shannon's entropy}
\label{Shannonentropy}

\begin{figure}[!htb]
    \centerline{\epsfxsize=9.cm \epsffile{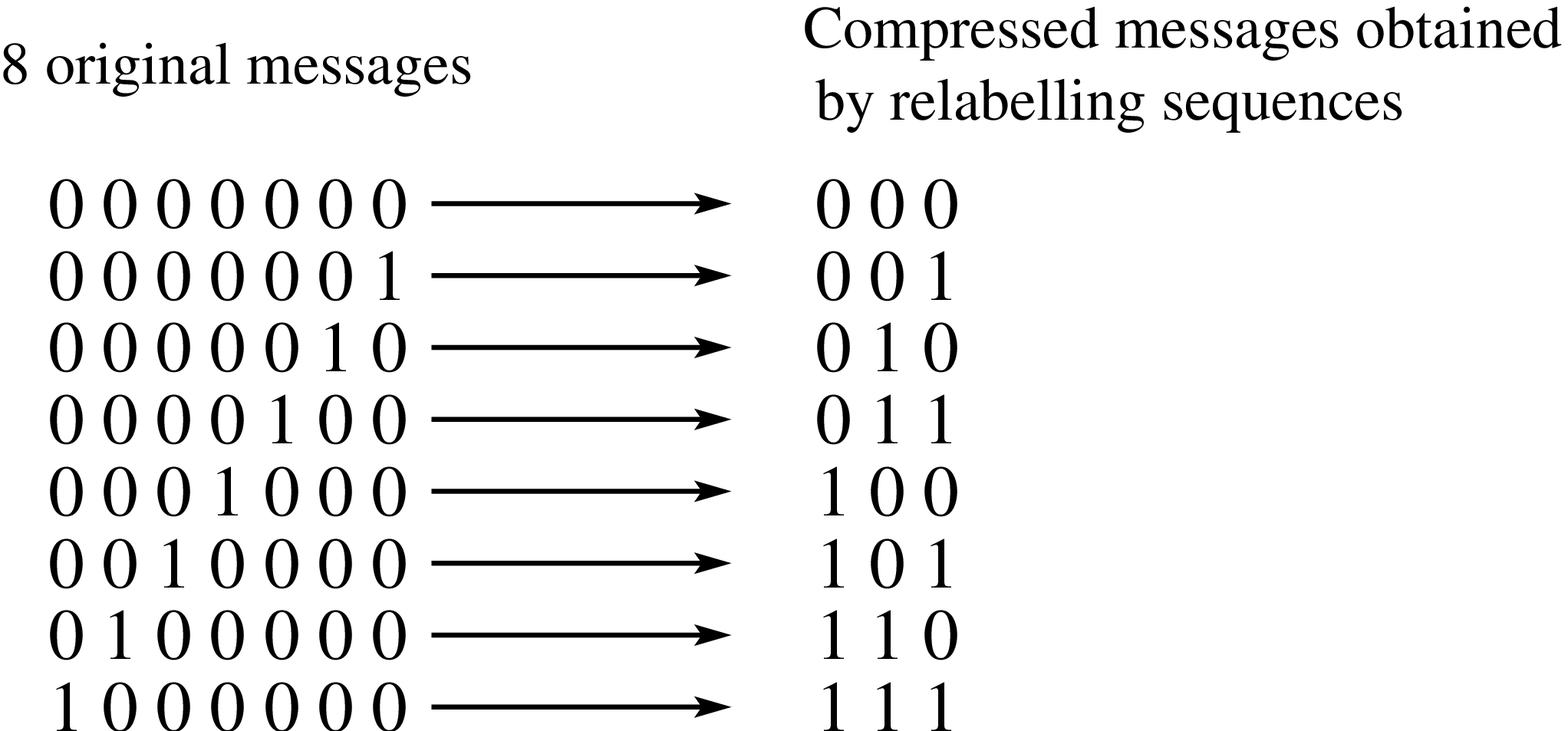}}
    \vspace*{0.15cm}
    \caption{The idea behind classical data compression. The most likely sequences are
    relabeled using fewer bits while rare sequences are discarded. The smaller number
    of bits still allows the reconstruction of the original sequences with very high
    probability. }
    \label{Shannon}
\end{figure}

In $1948$ Shannon developed a rigorous framework for the
description of information and derived an expression for the
information content of the message which indeed depends on the
probability of each letter occurring and results in the Shannon
entropy. We will illustrate Shannon's reasoning in the context of
the example above. Shannon invoked the law of large numbers and
stated that, if the message is composed of $N$ letters where $N$
is very large, then the $typical$ messages will be composed of
$Np_{1}$ 1's and $Np_{0}$ 0's. For simplicity, we assume that $N$
is 8 and that $p_1$ and $p_0$ are $\frac{1}{8}$ and $\frac{7}{8}$
respectively. In this case the typical messages are the $8$
possible sequences composed of 8 binary digits of which only one
is equal to $1$ (see left side of figure \ref{Shannon}).

As the length of the message increases (i.e. $N$ gets large) the probability of
getting a message which is all 1's or any other message that differs
significantly from a typical sequence is negligible so that we can safely
ignore them. But how many distinct typical messages are there? In the previous
example the answer was clear: just $8$. In the general case one has to find in
how many ways the $Np_{1}$ 1's can be arranged in a sequence of N letters?
Simple combinatorics tells us that the number of distinct typical messages is
\begin{equation}
    \left(\begin{array}{c} N \\ N p_1 \end{array}\right) = \frac{N!}{(Np_{1})!(Np_{0})!}
\end{equation}
and they are all equally likely to occur. Therefore, we can label each of these
possible messages by a binary number. If that is done, the number of binary
digits $I$ we need to label each typical message is equal to
$log_{2}\frac{N!}{Np_{1}!Np_{0}!}$. In the example above each of the 8 typical
message can be labeled by a binary number composed by $I = log_{2}8 = 3$ digits
(see figure \ref{Shannon}). It therefore makes sense that the number $I$ is
also the number of bits encoded in the message, because Alice can unambiguously
identify the content of each typical message if Bob sends her the corresponding
binary number, provided they share the background knowledge on the labeling of
the typical messages. All other letters in the original message are really
redundant and do not add any information! When the message is very long almost
any message is a typical one. Therefore, Alice can reconstruct with arbitrary
precision the original $N$ bits message Bob wanted to send her just by
receiving $I$ bits. In the example above, Alice can compress an 8 bits message
down to 3 bits. Though, the efficiency of this procedure is limited when the
message is only 8 letters long, because the approximation of considering only
typical sequences is not that good. We leave to the reader to show that the
number of bits $I$ contained in a large $N$-letter message can in general be
written, after using Stirling's formula, as

\begin{equation}
    I = -N(p_1 log p_1 + p_0 log p_0) \; .
    \label{Shannon example}
\end{equation}
If we plug the numbers $\frac{1}{8}$ and $\frac{7}{8}$ for $p_0$ and $p_1$
respectively in equation \ref {Shannon example}, we find that the information
content per symbol $\frac{I}{N}$ when N is very large is approximately $0.5436$
bits. On the other hand, when the binary letters 1 and 0 appear with equal
probabilities, then compression is not possible, i.e. the message has no
redundancy and each letter of the message contains one full bit of information
per symbol. These results match nicely the intuitive arguments given above.

Equation \ref {Shannon example} can easily be generalized to an
alphabet of $n$ letters $\rho_{i}$ each occurring with
probabilities $p_{i}$. In this case, the average information in
bits transmitted per symbol in a message composed of a large
number $N$ of letters is given by the Shannon entropy:

\begin{equation}
    \frac{I}{N} = H\{p_i\} = -\sum_{i=1}^n p_i log p_i \; .
    \label{Shannon entropy}
\end{equation}

We remark that the information content of a complicated classical
system composed of a large number $N$ of subsystems each of which
can be in any of $n$ states occurring with probabilities $p_i$
is given by $N \times H\{p_i\}$.

\subsubsection{Boltzmann versus Shannon entropy}
\label{BoltzmanversusShannon}

The mathematical form of the Shannon entropy $H$ differs only by a
constant from the entropy formula derived by Boltzmann after
counting how many ways are there to assemble a particular
arrangement of matter and energy in a physical system.

\begin{equation}
    S=-kln2\sum_{i=1}^n p_i log p_i \; .
    \label{Boltzmann entropy}
\end{equation}

To convert one bit of classical information in units of
thermodynamical entropy we just need to multiply by $kln2$. By
Landauer's erasure principle, the entropy so obtained is the
amount of thermodynamical entropy you will generate in erasing the
bit of information.

Boltzmann statistical interpretation of entropy helps us to understand the
origin of equation \ref{total entropy}. Consider our familiar example of the
binary device in which the molecule can be on either side of the partition with
equal probabilities. An observer who has no extra knowledge will use
Boltzmann's formula and work out that the entropy is $kln2$. What about an
observer who has 1 extra bit of information on the position of the molecule? He
will use the Boltzmann's formula again, but this time he will use the values
$1$ and $0$ for the probabilities, because he knows on which side the molecule
is. After plugging these numbers in equation \ref{Boltzmann entropy}, he will
conclude that the entropy of the system is $0$ in agreement with the result
obtained if we use equation \ref{total entropy}. The acquisition of information
about the state of a system changes its entropy simply because the entropy is a
measure of our ignorance of the state of the system as transparent from
Boltzmann's analysis.

\subsection{Sending classical information through a noisy classical channel}
\label{Sendingclassicalinformationthroughnoisyclassicalchannel}

In the previous section, we found that the Shannon entropy
measures the information content in bits of an arbitrary message
whose letters are encoded in classical objects. Throughout our
discussion, we made an important assumption: that the message is
encoded and transmitted to the recipient without errors. It is
obvious that this situation is quite unrealistic. In realistic
scenarios communication errors are unavoidable. To the physicist
eyes, the origin of noise in communication can be traced all the
way down to the unavoidable interaction between the environment
and the physical systems in which each letter is encoded. The
errors caused by the noise in the communication channel cannot be
eliminated completely. However, one hopes to devise a strategy
that enables the recipient of the message to detect and
subsequently correct the errors, without having to go all the way
to the sender to check the original message. This procedure is
sometimes referred to as {\em coding} the original message.

\subsubsection{Coding a classical message: an example}
\label{Codingaclassicalmessage:anexample}

For example, imagine that Bob wants to send to Alice a $1$ bit message encoded
in the state of a classical binary device in which a particle can be on the
left hand side (encode a $0$) or the right hand side (encode a $1$) of a finite
potential barrier. Unfortunately, the system is noisy and there is a
probability $\frac{1}{100}$ for the binary letter to flip (i.e. $1 \rightarrow
0$ or $0 \rightarrow 1$). For example, a thermal fluctuation induced by the
environment may cause the particle in the encoding device to overcome the
potential barrier and go from the left hand side to the right hand side. Alice,
who is not aware of this change, will therefore think that Bob attempted to
send a $1$ and not a $0$. This event occurs with $1\%$ probability so it is not
that rare after all. On the other hand, the (joint) probability that two such
errors occur in the same message is only $0.01\%$ ($\frac{1}{100} \times
\frac{1}{100}$). Alice and Bob decide to ignore the unlikely event of two
errors happening in one encoding but they still want to protect their message
against single errors. How can they achieve this?

One strategy is to add extra digits to the original message and
dilute the information contained in it among {\em all} the binary
digits available in the extended message. Here is an example.
Alice and Bob add two extra digits. Now their message is composed
of $3$ binary digits, but they still want to get across only one
bit of information. So they agree that Alice will read a $1$
whenever she receives the sequence $111$ and a $0$ when she
receives $000$.

The reader can see that this encoding ensures safer communication, because the
worst that can happen is that Alice receives a message in which not all the
digits are either 0s or 1s, for example $101$. But that is not big deal. In
this case the original message was clearly a $111$, because we have allowed for
single errors only. Under this assumption, any original message of the form
$000$ can never get transformed in $101$ because that requires flipping at
least two bits.

This strategy protects the message from single errors and therefore ensures
that the error rate in the communication is reduced down to 0.01\% (the
probability of double errors). By simply adding other two extra bits to the
encoded message Bob can protect the message against double errors and reduce
the error rate of two orders of magnitude (ie the probability of triple
errors). Quite obviously one can make the error rate as small as possible but
at the price of decreasing the ratio of $\frac{\mbox{bits
transmitted}}{\mbox{binary letters employed}}$. Is it possible to achieve a
finite ratio $\frac{\mbox{bits transmitted}}{\mbox{binary letters employed}}$
and an arbitrarily small error rate in the decoded messages? We will address
this question, that has been first answered by Shannon, in the next section.

\subsubsection{The capacity of a noisy classical channel via Landauer's principle}

Maybe surprisingly, one can indeed bring the error rate in the received message
in communication arbitrary close to zero, provided that the actual message of
length $N$ bits is "coded" in a much longer message of size $N_C$ bits. The
actual construction of efficient strategies to code a message is a task that
requires a lot of ingenuity, but is not what we are after. Our concern here is
to answer the following more fundamental question:

{\it Given that the probability of error is $q$, what is the largest number of
bits $N$ that we can transmit reliably through a noisy channel after encoding
them in a larger message of size $N_C$ bits?}

In other words we want a bound on the classical information
capacity of a noisy channel. We start by remarking that if the
coded message is composed of $N_C$ bits, then the average number
of errors will be $qN_C$. If we let the size of the message be
very large, the probability of getting a number of errors
different from the average value becomes vanishing small. In the
asymptotic limit one will expect exactly $qN_C$ bits to be
affected by errors in the $N_C$ bits message. However, there are
many ways in which $qN_C$ errors can be distributed in the $N_C$
bits of the original message. In fact, we worked out the exact
number in the section on the Shannon entropy and it is given by

\begin{equation}
    \mbox{number of ways the errors can be distributed} =
    \left(\begin{array}{c} N_C \\ q N_C \end{array} \right) \; .
    \label{errors}
\end{equation}
The problem there was slightly different, but after rephrasing the
argument a bit we can conclude that in order to specify how the
$qN_C$ errors are distributed among the $N_C$ message bits you
need $n$ bits of information, where $n$ is given by :

\begin{eqnarray}
    n &=& \log \left(\begin{array}{c} N_C
    \\ q N_C \end{array} \right) \cong -N_C[q log q + (1-q) log (1-q)] \nonumber\\
    & =& N_C H(q).
    \label{extra bits}
\end{eqnarray}

The reader should convince himself that equation \ref{extra bits}
can be derived following the same steps that led us to equation
\ref{Shannon example}. One just needs to rename the variables.

The short calculation above may inspire the following idea. Bob can send only
$N_C$ bits in total and he knows that he needs $N_C H(q)$ bits to specify the
position of the errors. All he has to do, then, is to allocate $N_C H(q)$
binary digits to store the information on the position of the errors. At that
point the remaining $N_C-N_C H(q)$ binary digits will be fully available for
safe communication. Unfortunately, Bob cannot implement this idea directly
because it requires him to know, in advance, which letters of the message are
going to be affected by errors. But the errors are random and they would occur
even in the letters that supposedly store information on their positions! But
there is something to be learned from this suggestion anyway.

Suppose, instead, that Bob had diluted the information he wants to transmit
among all the letters of the message as shown in the last section. When Alice
receives the string of binary digits and she deciphers the message, she gains
knowledge of the actual message, but also the information necessary to extract
the message from all the digits. This extra amount of information is implicitly
provided by the coding technique and it is also diluted among all the letters
in the message. To see this point more clearly, let us use Landauer's principle
and ask how much entropy Alice generates when she decides to erase the message
sent by Bob. For simplicity, let us stick to our simple example where Bob sends
$3$ bits to effectively transmit only a $1$ bit message. In order to erase the
information sent by Bob, Alice has to reset to zero the three classical binary
devices sent by Bob and that generates an amount of entropy not less than
$3kln2$, by Landauer's principle. But, Alice has effectively acquired only 1
bit of information corresponding to $kln2$ of entropy. So why did she have to
generate that extra amount of entropy equal to $2kln2$? Those extra 2 bits of
information that she is erasing must have been implicitly used to identify the
errors and separate them from the real message. In general, when Alice receives
the string of $N_C$ binary devices and she erases it, the minimum amount of
entropy that she generates is equal to $N_C \times kln2$. Now we can figure out
how much of that entropy needs to be wasted to extract the real message from
these (redundant) string of binary digits. No matter how sophisticated Bob's
coding was, there is no way that Alice could isolate the errors without using
at least $N_C H(q)$ bits of information. In fact, even if she can compress the
errors in a block of digits and concentrate the message in the remaining block
she would still need at least $N_C H(q)$ binary digits for the errors. Note
that we are by no means proving that she will be able to achieve this
efficiency, but only that she will compress the errors in a block of at least
$N_C H(q)$ binary letters. But, if Alice and Bob could device such a strategy,
something much more sophisticated than the naive idea suggested above, then
they would really have $N_C - N_C H(q)$ bits available for error free
communication. That means that there is an upper bound on the information
capacity of any classical noisy channel given by

\begin {equation}
    N = N_C(1 - H(q))
\end{equation}
where $N$ is the size of the message effectively transmitted,
$N_C$ is the size of the (larger) coded message and $q$ is the
probability that each bit will flip under the effect of the noise.
The rigorous proof that this bound is indeed achievable was given
by Shannon (see textbooks such as \cite{Cover T 91}). The reader
interested in more details can consult the Feynman lectures on
computation on which this short treatment was based
\cite{Feynmanbook}.

The problem of the noisy channel concludes our survey of classical information
encoded in classical systems. If you have a look at the map of this paper you
will see that we have gone through one of the 4 columns of topics shown
pictorially in figure \ref{structure}. The rest of this paper will deal with
topics that require a grasp of the basic principles and mathematical methods of
quantum mechanics. The next section is a quick recap that should be of help to
those with a more limited background. If the reader feels confident in the use
of the basics of quantum mechanics, the density operator and tensor products,
then he can just skip this part and move on to the next section.

\section{A crash course on quantum mechanics}
\label{Crashcourse}

At the end of our discussion on the Maxwell's demon paradox, we started putting
forward the idea that the information we have on the state of a classical
system contributes to define the state itself. In this section we will push our
arguments even further and investigate the role that the concept of information
plays in the basic formalism of quantum mechanics.

\subsection{To be or to know}
\label{Tobe}

The quantum state of a physical system is usually represented mathematically by
a vector $\ket{\psi}$ or a matrix ${\hat \rho}$ in a complex vector space
called the Hilbert space \cite{Preskill notes,Plenio notes,Mabuchi notes,Peres
95}. We will explain the rules and the reasoning behind this representation in
the next sections by considering two-level quantum systems as an easy example
that displays most of the features of the general case.

But, first of all, {\it what do the mathematical symbols exactly
represent?} In this article, we take the pragmatic point of view
that {\it what is being represented is not the quantum system
itself but rather the information that we have about its
preparation procedure}. As an example that illustrates this point,
we consider the process by which an atom prepared in an arbitrary
superposition of energy eigenstates collapses into only one of the
eigenstates after the measurement is done. This process seems to
happen instantaneously unlike the ordinary time evolution of
quantum states. Generations of physicists have been puzzled by
this fact and have searched for the physical mechanism which
causes the collapse of the wave function. However, if we consider
the wave-function to represent only the information that we
possess about the state of the quantum system, we will definitely
expect it to change discontinuously after the measurement has
taken place, because our knowledge has suddenly increased. Not
everybody is satisfied with this view. Some people think that
physical theories should deal with objective properties of Nature,
with what is {\em really} out there and avoid subjectivism. It is
difficult to assess the validity of these arguments entirely on
philosophical grounds. To our knowledge there are no experiments
that provide compelling evidence in favor of any of the existing
interpretational frameworks. Therefore we will adopt what we feel
is the easiest way out of the problem and explain the rules for
representing mathematically our knowledge of the preparation
procedure of an arbitrary quantum state \cite{Fuchs P 00}.

\subsection{Pure states and complete knowledge}
\label{purestates}

\subsubsection{Pure states of a single system}
\label{Purestatesofasinglesystem}

We start by considering how to proceed when we have complete knowledge on the
preparation procedure of a single quantum system. In this simpler case, we say
that the state of the quantum system is {\em pure} and we represent our
complete knowledge of its preparation procedure as a vector in a complex vector
space. As an example, consider two non-orthogonal states of a two-level atom
$\ket{\psi_1}$ and $\ket{\psi_0}$. These states are arbitrary superpositions of
the {\em two} energy eigenstates. In the next few lines, we show how to write
them as {\em two} $2$-dimensional vectors

\begin{eqnarray}
    \ket{\psi_1} &=& \frac{2}{\sqrt {5}}\ket{0}
    + \frac{1}{\sqrt{5}}\ket{1} \;\;.
    \nonumber\\
    &=&
        \frac{2}{\sqrt {5}}\left( \begin{array}{c}  1 \\ 0 \end{array} \right) +
        \frac{1}{\sqrt {5}}\left( \begin{array}{c}  0 \\ 1 \end{array} \right) \;\;.
    \nonumber\\
    &=&
        \frac{1}{\sqrt {5}}\left( \begin{array}{c}  2 \\ 1 \end{array} \right)\;\; .
\label{psi1}
\end{eqnarray}
\begin{eqnarray}
    \ket{\psi_0} &=& \frac{1}{\sqrt {2}}\ket{0}
    + \frac{1}{\sqrt{2}}\ket{1} \;\;.
    \nonumber\\
    &=&
        \frac{1}{\sqrt {2}}\left( \begin{array}{c}  1 \\ 1 \end{array} \right)\;\; .
\label{psi0}
\end{eqnarray}
The rule used above to convert from Dirac to matrix notation is to write the
energy eigenstates $\ket{0}$ and $\ket{1}$, as the column vectors
$\left(\begin{array}{c}  1 \\ 0 \end{array} \right)$ and
$\left(\begin{array}{c}  0 \\ 1 \end{array} \right)$, respectively. There is
nothing {\em mystical} behind the choice of this correspondence. One could have
also chosen the basis vectors $\frac{1}{\sqrt{2}}\left(\begin{array}{c}  1 \\ 1
\end{array} \right)$ and $\frac{1}{\sqrt{2}}\left(\begin{array}{c} -1 \\ 1
\end{array} \right)$, instead. What is important is that the two vectors are
$orthogonal$ and normalized so that they can faithfully represent the important
$experimental$ property that the two states $\ket{0}$ and $\ket{1}$ are
orthogonal and can be perfectly $distinguished$ in a measurement. The important
point to observe in the choice of the basis in which to represent your
state-vectors is that of consistency. Every physical quantity has to be
represented in the same basis when you bring them together in computations. If
one has used different bases for representation, then one has to rotate them
into one standard basis using unitary transformations. This rotation can be
expressed mathematically as $2 \times 2$ unitary matrix $U$. A unitary matrix
is defined by the requirement that $UU^{\dagger}=U^{\dagger}U={\bf 1}$. Given a
set of quantities in one basis then upon rewriting them in another basis, the
predictions for all physically observable quantities have to remain the same.
This essentially requires that the mathematical expressions that are used to
express these observable quantities have to be invariant under unitary
transformations. We will see examples of this soon.

Above we have seen examples for orthogonal states (namely the basis states
$\ket{0}$ and $\ket{1}$, as the column vectors $\left(\begin{array}{c} 1
\\ 0 \end{array} \right)$ and $\left(\begin{array}{c}  0 \\ 1 \end{array}
\right)$ ). In general two quantum states will be neither
orthogonal nor parallel such as for example the states
$\ket{\psi_0}$ and $\ket{\psi_1}$. To quantify the angle between
two vectors $\ket{\psi_i}$ and $\ket{\psi_j}$ we introduce the
complex scalar product. For complex vectors with two components it
is given by
\begin{eqnarray}
    \bra{\psi_j}\psi_i\rangle &=& ({a_j}^* \bra{0} + {b_j}^* \bra{1})(a_i\ket{0} + b_i\ket{1}) \;\;.
    \nonumber\\
    &=& \left( \begin{array}{c c}
    {a_j}^* & {b_j}^*
    \end{array} \right) \left( \begin{array}{c} a_i \\ b_i \end{array} \right)
         \;\;.
    \nonumber\\
    &=&
    {a_j}^* a_i + {b_j}^* b_i \;\; .
\label{dot product}
\end{eqnarray}
Note that the components of the first vector have to be complex conjugated, but
apart from that the complex scalar product behaves just as the ordinary real
scalar product. One nice property of the scalar product is the fact that it is
invariant under unitary transformations, just as you would expect for a
quantity that measures the angle between two state vectors.

\subsubsection{Operators and probabilities for a single system}

In our new language of state vectors, the dot product
$\bra{\psi_i}\psi_j\rangle$ is analogous to the overlap integral between two
wave-functions $\psi_i(x)$ and $\psi_j(x)$, that is usually encountered in
introductory courses of quantum mechanics. The reader may recall that the {\em
squared} result of the overlap integral, write as
${|\bra{\psi_i}\psi_j\rangle|}^{2}$, can be interpreted as the probability of
projecting the quantum state $\ket{\psi_i}$ on the eigenstate $\ket{\psi_j}$ of
an appropriate observable after the measurement is performed.

Now we would like to represent this projection mathematically by a projection
operator denoted by $\projector{\psi}$. This projector is simply a matrix that
maps all the vectors onto the vector corresponding to $\ket{\psi_j}$, apart
from a normalization constant. The recipe to construct the matrix
representation of $\projector{\psi}$ is to multiply the column vector
$|\psi\rangle$ times the row vector $\bra{\psi}$ as shown below:

\begin{eqnarray}
    \projector{\psi} &=& (a\ket{0} + b\ket{1})(a^*\bra{0} + b^*\bra{1}) \;\;.
    \nonumber\\[0.2cm]
    &=&
        \begin{array}{c} \\  \left( \begin{array}{c} a \\ b \end{array} \right)
        \\
        \end{array} \hspace*{-0.2cm}
        \begin{array}{c} \left( \begin{array}{c c} {a}^* & {b}^* \end{array} \right) \\ \\ \\ \end{array}
    \nonumber\\[0.1cm]
    &=&
    \left( \begin{array}{c c}
    |a|^2 & a {b}^* \\
    {a}^*b & |b|^2
    \end{array} \right) \;\; .
\label{Mabuchi5.3}
\end{eqnarray}
For example, the reader can easily construct the matrix representing the
projector $\projector{1}$ and check that when it operates on the state
$\ket{\psi_0}$ in equation $\ref{psi0}$ we indeed obtain the excited state
$\ket{1}$ apart from a normalization constant. Furthermore, the probability of
finding the state $|\psi\rangle$ in a measurement of a system originally in the
quantum state $\ket{\phi}$ is given by

\begin {equation}
    Prob_{|\psi\rangle} = \langle\phi| (|\psi\rangle\langle\psi|)
    |\phi\rangle =
    tr\{|\psi\rangle\langle \psi|
    |\phi\rangle\langle \phi|\} \label{prob proj}
\end{equation}
where $tr$ denotes the trace which is the sum of the diagonal elements of a
matrix, a concept that is invariant under unitary transformations. The reader
can easily check that Eq. \ref{prob proj} is true by explicitly constructing
the matrices $|\psi\rangle\langle \psi|$ and $|\phi\rangle\langle \phi|$ (see
equation \ref{Mabuchi5.3}), multiplying them, take the trace, and verify that
the result is indeed equal to $|\bra{\phi}\psi\rangle|^{2}$, calculated after
squaring the result of equation \ref{dot product}. Once this is done it is easy
to write the expectation value of any observable  whose eigenvalues are the
real numbers $\{e_i\}$ and its eigenstates are the vectors $\{\ket{e_i}\}$. In
fact, if we label the probability of projecting on the eigenstate $\ket{e_i}$
as $Prob_{\ket{e_i}}$ and we make use of equation \ref{prob proj}, we can
indeed write the expectation value for any observable ${\hat O}$ of the two
level system in a given state $|\phi\rangle$ as

\begin{eqnarray}
    \langle {\hat O} \rangle_{|\phi\rangle} &=& e_0 Prob_{|e_0\rangle} + e_1
    Prob_{|e_1\rangle}\;\;. \nonumber\\ &=& e_0 tr\{|e_0\rangle\langle e_0|
    |\phi\rangle\langle \phi|\}+ e_1 tr\{|e_1\rangle\langle e_1|
    |\phi\rangle\langle \phi|\}\;\;. \nonumber\\ &=& tr\{(e_0 |e_0\rangle\langle
    e_0|+ e_1 |e_1\rangle\langle e_1|) |\phi\rangle\langle \phi|\}\;. \label{1}
\end{eqnarray}
The expression above can be tided up a bit by defining the observable ${\hat
O}$ as the matrix

\begin{equation}
    {\hat O} = e_0 |e_0\rangle\langle e_0|+ e_1 |e_1\rangle\langle e_1|
    \label{O}\; .
\end{equation}
Note that in order to use the projectors to calculate probabilities as in
equation \ref{1}, we have to demand that the sum of the matrices representing
the projectors must be the unity matrix. For a two dimensional vector space
this means that $\projector{0}+\projector{1}=1 $. This condition ensures that
the sum of the probabilities obtained using equation \ref{1} is equal to $1$.
Once we check this important property of the projectors we can use equation
\ref{O} to construct the matrix representation of any observable. For example,
the reader can check that the energy observable ${\hat E}$ can be written using
the basis $\left(\begin{array}{c} 1 \\ 0
\end{array} \right)$ and $\left(\begin{array}{c}  0 \\ 1
\end{array} \right)$ in the form:

\begin{eqnarray}
    {\hat E} &=& e_0 \projector{e_0} + e_1 \projector{e_1} \;\;.
    \nonumber\\
    &=&
        e_0 \left( \begin{array}{c c}
        1 & 0 \\
        0 & 0
        \end{array} \right)+
        e_1 \left( \begin{array}{c c}
    0 & 0 \\
    0 & 1
    \end{array} \right) \;\;.
    \nonumber\\
    &=&
    \left( \begin{array}{c c}
    e_0 & 0 \\
    0 & e_1
    \end{array} \right) \;\; .
    \label{energy}
\end{eqnarray}
Note that the energy operator is diagonal in this basis because these basis
vectors were originally chosen as the energy eigenvectors! However, the
prescription given in equation \ref{O} to represent any observable ${\hat O}$
ensures that the resulting matrix is Hermitian because the projectors
themselves are Hermitian. A matrix is said to be Hermitian if all its entries
that are symmetrical with respect to the principal diagonal are complex
conjugate of each other (see equation \ref{Mabuchi5.3}). The fact that the
matrix ${\hat O}$ is Hermitian ensures that its eigenvectors are orthogonal and
the corresponding eigenvalues are real. This means that the possible "output
states" after the measurement are distinguishable and the corresponding results
are real numbers. Once you accept equation \ref{O}, you can immediately write
equation \ref{1} simply as

\begin{equation}
    \langle {\hat O} \rangle = tr\{{\hat O}|\psi_i\rangle\langle \psi_i|\}
    \label{expectO}
\end{equation}
This completes our quick survey of the rules to represent the
arbitrary state of a single two level quantum system. The main
motivation to adopt these rules is dictated by their ability to
correctly predict experimental results.

\subsubsection{Non-orthogonality and inaccessible information}
\label{nonorthogonal}

We would like to expand a little bit on the important concept of {\it
non-distinguishibility} between two quantum states. By this we mean the
following. Suppose that you are given two two-level atoms in states
$\ket{\psi_0}$ and $\ket{\psi_1}$ respectively (see equations \ref{psi0} and
\ref{psi1}) and you are asked to work out which particle is in state
$\ket{\psi_1}$ and which in state $\ket{\psi_0}$. The two states are said to be
non-distinguishable if you will never be able to achieve this task without the
possibility of a wrong answer and if you are given only one system and
irrespective of the observable you to measure. For example you could decide to
measure the energy of the two atoms. After using equation \ref{prob proj}or
just by inspection, you can verify that the probability of finding the atom in
the excited state if it was in state $\ket{\psi_0}$ before the measurement is
equal to $\frac{1}{5}$. On the other hand, you can also check that the
probability of finding the atom in the excited state if it was in state
$\ket{\psi_1}$ before the measurement is also non-vanishing and in fact equal
to $\frac{1}{2}$. Now, suppose that you perform the measurement and you find
that the atom is indeed in the excited state. At this point, you still cannot
unambiguously decide whether the atom had been prepared in state $\ket{\psi_0}$
or $\ket{\psi_1}$ before the measurement took place. In fact, {\it by measuring
any other observable only once you will never be able to distinguish between
two non-orthogonal states with certainty}.

This situation is somehow surprising because the two non-orthogonal states are
generated by different preparation procedures. Information was invested to
prepare the two states, but when we try to recover it with a single measurement
we fail. The information on the superposition of states in which the system was
prepared remains not accessible to us in a single measurement.

It is sometimes argued that we therefore have to assume that a single quantum
mechanical measurement does not give us any information. This viewpoint is,
however, wrong. Consider the situation above again, where we either have the
state $|\psi_0\rangle$ or the state $|\psi_1\rangle$ with a priori
probabilities $1/2$ each. If we find in a measurement the excited state of the
atom, then it would be a fair guess to say that it is more likely that the
system was in state $|\psi_1\rangle$ because this state has the higher
probability to yield the excited state in a measurement of the energy.
Therefore the a posteriori probability distribution for the two states has
changed, and therefore we have gained knowledge as we have reduced our
uncertainty about the identity of the quantum state.

The non-distinguishability of non-orthogonal quantum states is an
important aspect of quantum mechanics and will be encountered
again several times in the remainder of this article.

\subsubsection{Two 2-level quantum systems in a joint pure state}

We have gained a good grasp of the properties of an isolated two-level quantum
system. We are now going to study how the joint quantum state of two such
systems (say a pair of two level atoms) is represented mathematically. The
generalization is straightforward. We initially concentrate on the situation
when our knowledge of the preparation procedure of the joint state is complete,
i.e. when the joint system is in a pure state. The reader who is not very
familiar with quantum mechanics may wonder why we have to include this section
altogether. At the end of the day, according to classical intuition, the state
of a joint quantum system comprised of two subsystems $A$ and $B$ can be given
by simply providing, at any time, the state of each of the subsystems $A$ and
$B$ independently. This reasonable conclusion turns out to be wrong in many
cases! Let us see why.

We first consider one of the most intuitive examples of joint state of the two
atoms: the case in which atom $A$ is in its excited state $\ket{1}_A$ and atom
$B$ in its ground state $\ket{0}_B$, where the subscript labels the atoms and
the binary number their states. In this case, the joint state of the two atoms
$\ket{\psi_{AB}}$ can be fully described by stating the state of each atom
individually so we write $\ket{\psi_{AB}}$ down symbolically as $\ket{1}_A$
$\ket{0}_B$. We call this state a {\em product state}. We now decide to
represent the joint state $\ket{1}_A \ket{0}_B$ of the two atoms as a vector in
an enlarged Hilbert space whose dimensionality is no longer $2$ as for a single
atom but it is $2 \times 2 = 4$. The vector representation of $\ket{1}_A
\ket{0}_B$ is constructed as shown below:

\begin{eqnarray}
    \ket{\psi_{AB}}&=& \ket{1}_A \ket{0}_B
    \nonumber\\
    &=&
        \left( \begin{array}{c} 0 \\ 1 \end{array} \right)\otimes
        \left( \begin{array}{c}
         1 \\ 0
    \end{array} \right)
    \nonumber\\
    &:=&
    \left( \begin{array}{c}
    0 \times 1 \\
    0 \times 0 \\
    1 \times 1  \\
    1 \times 0 \end{array} \right)
\nonumber\\
    &=&
    \left( \begin{array}{c} 0  \\ 0  \\ 1 \\ 0 \end{array} \right) \;\; .
\label{Tensor1}
\end{eqnarray}
Equation \ref{Tensor1} defines the so called tensor product
between two vectors belonging to two different Hilbert spaces, one
used to represent the state of atom A and the other for atom B.
For the readers who have never seen the symbol $\otimes$ we write
down a more general case involving the two vectors $\ket{\psi_A}$
with coefficients $a$ and $b$ and $\ket{\psi_B}$ with coefficients
$c$ and $d$:

\begin{eqnarray}
    \ket{\psi_{AB}}&=& \ket{\psi_A}\ket{\psi_B}
    \nonumber\\
    &=&
        \left( \begin{array}{c} a \\ b \end{array} \right)
        \left( \begin{array}{c}
         c \\ d
    \end{array} \right)
    \nonumber\\
    &:=&
    \left( \begin{array}{c} ac \\ ad \\ bc  \\ bd \end{array} \right) \;\; .
\label{Tensor2}
\end{eqnarray}
The case of tensor product between $n$ dimensional vectors is a
simple generalization of the rule of multiplying component-wise as
above \cite{Plenio notes}. Using equation \ref{Tensor2} the reader
can work out the vector representation of the following states:

\begin{equation}
    \ket{0}_A\ket{0}_B \longrightarrow \left( \begin{array}{c} 1 \\ 0 \\ 0 \\ 0
    \end{array} \right) , \label{arr1}
\end{equation}
\begin{equation}
    \ket{0}_A\ket{1}_B \longrightarrow \left( \begin{array}{c} 0 \\
    1 \\ 0 \\ 0 \end{array} \right) ,
    \label{arr2}
\end{equation}
\begin{equation}
    \ket{1}_A\ket{1}_B \longrightarrow \left( \begin{array}{c} 0 \\
    0 \\ 0 \\ 1 \end{array} \right).
    \label{arr3}
\end{equation}
A trick to write the states above as vectors without explicitly
performing the calculation in equation \ref{Tensor1} is the
following. First, read the two digits inside $\ket{...}\ket{...}$
as two digits binary numbers (for example read $\ket{0}\ket{1}$ as
1), and add $1$ to get the resulting number $n$. Then place a $1$
in the $n^{th}$ entry of the column vector and 0s in all the
others. The four states-vectors in equations \ref{Tensor1},
\ref{arr1}, \ref{arr2} and \ref{arr3} are a complete set of
orthogonal basis vectors for our four-dimensional Hilbert space.
Therefore, any state $\ket{\psi_{AB}}$ of the form
$\ket{\psi_B}\ket{\psi_A}$ in equation \ref{Tensor2} can be
written as:

\begin{equation}
\ket{\psi_{AB}} = ac \ket{0}_A \ket{0}_B + ad \ket{0}_A \ket{1}_B + bc \ket{1}_A \ket{0}_B
+ bd \ket{1}_A \ket{1}_B.
\label{sup}
\end{equation}

where we wrote the vectors symbolically, in Dirac notation, to save paper. We
interpret the coefficients of each basis vector in terms of probability
amplitudes, as we did for single systems. For example, the modulus squared
$|ad|^{2}$ gives the probability of finding atom $A$ in its ground state and
atom $B$ in the excited state after an energy measurement. A question that
arises naturally after inspecting equation above is the following:

{\it What happens when I choose the coefficients of the
superposition in equation \ref{sup} in such a way that it is
impossible to find two vectors $\ket{\sigma}_A$ and
$\ket{\beta}_B$ that "factorize" the 4-dimensional vector
$\ket{\psi_{AB}}$ as in equation \ref{Tensor2}? Are these non
factorizable vectors a valid mathematical representation of
quantum states that you can actually prepare in the lab?}

\subsubsection{Bipartite Entanglement}
\label{BipartiteEntanglement}

The answer to the previous question is a definite yes. Before expanding on this
point, let us write an example of a non factorizable vector:

\begin{equation}
\ket{\psi_{AB}} = \frac{1}{\sqrt{2}} \ket{0}_A \ket{0}_B + \frac{1}{\sqrt{2}} \ket{1}_A \ket{1}_B.
\label{maxent}
\end{equation}
The vector above corresponds to the state for which there is equal probability
of finding both atoms in the excited state or both in the ground state. The
reader can perhaps make a few attempts to factorize this vector, but they are
all going to be unsuccessful. This vector, nonetheless, represents a perfectly
acceptable quantum state. In fact, according to the laws of quantum mechanics,
ANY vector in the enlarged Hilbert space is a valid physical state for the
joint system of the two atoms, independently of it being factorizable or not.
In fact, in section \ref{nqubits} we will show that for an $n$-partite system
most of the states are actually non factorizable. So these states are the norm
rather than the exception!

The existence of non-factorizable states is not too difficult to
appreciate mathematically, but it leads to some unexpected
conceptual conclusions. If the quantum state of a composite system
cannot be factorized than it is impossible to specify a pure state
of its constituent components. More strangely perhaps,
non-factorizable states, such as $\ket{\psi_{AB}}$ in equation
\ref{maxent} are pure states. This means that the corresponding
vectors are mathematical representations of our {\em complete}
knowledge of their preparation procedure. There is nothing more we
can {\em in principle} know about these composite quantum objects
than what we have written down, but nonetheless we still cannot
have full knowledge of the state of their subsystems. With
reference to the discussion following equation \ref{maxent}, we
conclude that in a non-factorizable state we have knowledge of the
correlation between measurements outcomes on atom $A$ and $B$ but
we cannot in principle identify a pure state with each of the
atoms $A$ and $B$ individually. This phenomena seemed very weird
to the fathers of quantum mechanics who introduced the name
entangled states to denote states whose corresponding vectors
cannot be factorized in the sense explained above. In section
\ref{Entanglement}, that is entirely devoted to this topic, we
will go beyond the dry mathematical notion of non-factorizability
and start exploring the physical properties that make entangled
states peculiar.  We will focus on possible applications of
these weird quantum objects in the lab. But before doing that, the
reader will have to swallow another few pages of definition and
rules because we have not explained yet how to construct and
manipulate operators acting on our enlarged Hilbert space.

\subsubsection{Operators and probabilities for two systems}

In this section, we generalize the discussion of projection operators and
observables given previously for single quantum systems to systems consisting
of two particles. The generalization to $n$-particle systems should then be
obvious. We start by asserting that the rules stated in equations \ref{prob
proj} and \ref{O} for single quantum systems are still valid with the only
exception that now observables and projector operators are represented by $4
\times 4$ matrices. Imagine that you want to write down the joint observable
${\hat O}_A \otimes {\hat O}_B$ where ${\hat O}_A$ and ${\hat O}_B$ are
possibly different observables acting respectively on the Hilbert space of
particle A and of particle B. The rule to write down the joint observable is
the following:

\begin{eqnarray}
    {\hat O}_{AB} &=& {\hat O}_{A} \otimes {\hat O}_{B} \;\;.
    \nonumber\\
    &=&
        \left( \begin{array}{c c}
        a_1 & b_1 \\
        c_1 & d_1
        \end{array} \right) \otimes \left( \begin{array}{c c}
        a_2 & b_2 \\
        c_2 & d_2
        \end{array} \right) \;\;.
    \nonumber\\
    &=&
    \left( \begin{array}{c c c c}
    a_1 a_2 & a_1 b_2 & b_1 a_2 & b_1 b_2 \\
    a_1 c_2 & a_1 d_2 & b_1 c_2 & b_1 d_2 \\
    c_1 a_2 & c_1 b_2 & d_1 a_2 & d_1 b_2 \\
    c_1 c_2 & c_1 d_2 & d_1 c_2 & d_1 d_2
\end{array} \right) \;\; .
    \label{en}
\end{eqnarray}
where the subscript $1$ denotes the operator on particle $A$ and the subscript
$2$ the operator on particle $B$. However, there are some observables ${\hat
O_{AB}}$ whose corresponding matrices cannot be factorized as in equation
\ref{en}. These matrices still represent acceptable observables provided that
they are Hermitian.

Furthermore, it is possible to construct projectors on any $4d$ vectors by
using the same principle illustrated in equation \ref{Mabuchi5.3}. For example,
the projector on the entangled state $\ket{\psi}_{AB}$ in equation \ref{maxent}
can be written as

\begin{eqnarray}
    \projector{\psi_{AB}}
    &=&
        \frac{1}{2} \left( \begin{array}{c} 1 \\ 0 \\ 0 \\ 1 \end{array} \right)
        \left( \begin{array}{c c c c}
    1 & 0 & 0 & 1
    \end{array} \right) \;\;.
    \nonumber\\
    &=&
    \frac{1}{2}\left( \begin{array}{c c c c}
    1 & 0 & 0 & 1 \\
    0 & 0 & 0 & 0 \\
    0 & 0 & 0 & 0 \\
    1 & 0 & 0 & 1
\end{array} \right) \;\; .
\label{projent}
\end{eqnarray}
Finally, suppose you are interested in knowing the probability of projecting
atom A on its ground state $\ket{0}_A$ and atom B onto its excited state
$\ket{1}_B$ after performing a measurement on the maximally correlated state
$\ket{\psi_{AB}}$ considered above. How do you proceed? The answer to this
question should be of guidance also for other cases, so we work it out in some
detail. The first thing you do is to construct the tensor product of the
matrices corresponding to the single particle projectors $\projector{0}$ and
$\projector{1}$ that project particle A onto its ground state and particle B on
its excited state:

\begin{eqnarray}
    \projector{0} \otimes \projector{1}
    &=&
\left( \begin{array}{c c} 1 & 0 \\
                    0 & 0
\end{array} \right) \otimes
\left( \begin{array}{c c}
    0 & 0 \\
    0 & 1
    \end{array} \right) \;\;.
    \nonumber\\
    &=&
    \left( \begin{array}{c c c c}
    0 & 0 & 0 & 0 \\
    0 & 1 & 0 & 0 \\
    0 & 0 & 0 & 0 \\
    0 & 0 & 0 & 0
\end{array} \right) \;\; .
\label{tensor projector}
\end{eqnarray}
Once you have worked out the matrix in equation \ref{tensor
projector} you can multiply it with the matrix found in equation
\ref{projent} and take the trace, as explained for single
particles in equation \ref{prob proj}. The result is 0, as
expected, since we have maximal correlations between the two atoms
in state $\ket{\psi_{AB}}$.

\subsection{Mixed states and incomplete knowledge}
\label{Pureversusmixed}

\subsubsection{Mixed states of a single two-level atom}
\label{Thecreation}

In this section, we explain how to represent mathematically the state of a
quantum system whose preparation procedure is not completely known to us. This
lack of knowledge may be caused by random errors in the apparatus that
generates our quantum systems or by fluctuations induced by the environment. In
these cases we say that the quantum system is in a {\em mixed state}. This can
be contrasted with the pure states considered in the previous sections for
which there was no lack of knowledge of the preparation procedure (i.e. the
quantum states were generated by a perfect machine whose output was completely
known to us). To some extent, by considering mixed states, we start dealing
with "real world quantum mechanics". We will build on the example introduced in
section \ref{Purestatesofasinglesystem} to make our treatment more accessible.

\begin{figure}[!htb]
    \centerline{\epsfxsize=8cm \epsffile{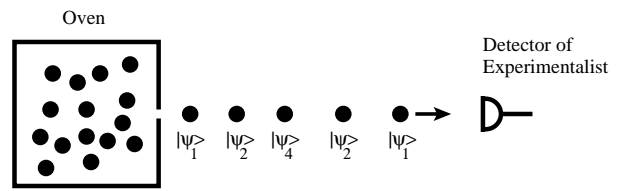}}
    \vspace*{0.15cm}
    \caption{An oven emits atomic two-level systems. The internal state of the
    system is randomly distributed. With probability $p_i$ the system
    is in the pure state $|\psi_i\rangle$. A person oblivious to this
    random distribution measures observable ${\hat A}$. What is the
    mean value that he obtains?}
    \label{experiment}
\end{figure}

An experimentalist needs to prepare two-level atoms in the state
$\ket{\psi_1}$ to be subsequently used in an experiment. He has at
his disposal an oven that generates atoms in the state
$\ket{\psi_1}$ with probability $p_1 = 95\%$ (see Fig. \ref{experiment}
for illustration).
In the remaining
$p_0 = 5\%$ of the cases the oven fails and generates atoms in a
different state $\ket{\psi_0}$. This preparation procedure is
pretty efficient, but of course still different from the ideal
case. The experimentalist collects the atoms, but {\it he does not
know for which of them the preparation has been successful}
because the experimental errors occur randomly in the oven.
Neither can he measure the atoms because he is scared of
perturbing their quantum state. The only thing he knows is the
{\it probability distribution of the two possible states}. The
experimentalist has to live with this uncertainty. However, he his
aware that, if he uses the states produced by the oven, his
experimental results are going to be different from the ones he
would have obtained had he used atoms in the state $\ket{\psi_1}$
exactly, because the oven occasionally outputs atoms in the
undesired state $\ket{\psi_0}$. He would like to find an easy way
to compute the measurement results in this situation so he asks a
theorist to help him modeling his experiments.
The first task the two have to face is to construct a mathematical object that
represents their {\em incomplete} knowledge of the preparation procedure.
Intuitively, it cannot be the vector $\ket{\psi_1}$ because of that $5\%$
probability of getting the state $\ket{\psi_0}$. The way the two approach the
problem is a good example of empirical reasoning, so it is worth exploring
their thought process in some detail. The theorist asks the experimentalist to
describe what he needs to do with these atoms and the two reach the conclusion
that what really matters to them are the expectation values of arbitrary
observables measured on the states generated by the oven. The theorist points
out that, after performing measurements on $N$ atoms, the experimentalist will
have used, approximately, $N p_1$ atoms in the state $|\psi_1\rangle$ and $N
p_0$ atoms in state $|\psi_0\rangle$. For each of the two states
$|\psi_i\rangle$ they would know how to calculate the expectation value for any
observable ${\hat A}$ that the experimentalist wants to measure. After using
equation \ref{O} the theorist rewrites the expectation value of the observable
${\hat A}$ on the state $|\psi_i\rangle$ as $tr\{{\hat A} |\psi_i\rangle\langle
\psi_i| \}$. The two are now only one step away from the result. What they need
to do is to average the two expectation values for the states $|\psi_1\rangle$
and $|\psi_0\rangle$ with the respective probabilities. The mean value observed
by the experimentalist is thus given by:

\begin{eqnarray}
    \langle {\hat A} \rangle &=&
    \sum_i p_i tr\{{\hat A} |\psi_i\rangle\langle \psi_i| \} \nonumber\\
    &=& tr\{{\hat A} \sum p_i |\psi_i\rangle\langle \psi_i| \}\;\; .
    \label{calculation}
\end{eqnarray}
The calculation above can be tided up a bit by defining the density operator
${\hat\rho}$ as

\begin{eqnarray}
    {\hat\rho} = \sum p_i |\psi_i\rangle\langle \psi_i| \;\; .
    \label{density operator}
\end{eqnarray}
Once this is done equation \ref{calculation} can be compactly written as

\begin{eqnarray}
    \langle {\hat A} \rangle = tr\{{\hat A} {\hat\rho} \} \;\; .
\label{expectation value}
\end{eqnarray}
A glance at these few lines of mathematics convinces the two
physicists that they have actually solved their problem. In fact
the density operator is the mathematical description of the
knowledge the two have about the quantum states prepared by the
oven. Equation \ref{expectation value}, on the other hand, tells
them exactly how to use their knowledge to compute the expectation
value of any operator.

Similarly, they can write down the probability of finding the system in any
state $\ket{\sigma}$ after a measurement by simply constructing the projector
$\projector{\sigma}$. After this, they just multiply it with the density
operator and take the trace (as in equation \ref{prob proj})

\begin{eqnarray}
    Prob_{\ket{\sigma}} = tr\{\projector{\sigma}{\hat\rho} \} \;\; .
\label{prob proj2}
\end{eqnarray}
Equation \ref{density operator} provides the recipe for constructing the
density matrix for the example above. We leave as an exercise to the reader to
show that the density operator representing the preparation procedure described
above can be written as

\begin{equation}
    {\hat \rho} =
          \left( \begin{array}{c c}
        0.785 & 0.405 \\
        0.405 & 0.215 \\
    \end{array} \right)
    \label{densoper}
\end{equation}
One can see that the trace of the density operator ${\hat \rho}$ in equation
\ref{densoper} is equal to $1$. This is not an accident but a distinctive
property of any density operator. You can easily check that by plugging the
unity matrix rather than the operator ${\hat O}$ in equation \ref{expectation
value}. The expectation value of the unity operator on any normalized vector
state is $1$ (i.e. the expectation value reduces to the dot product of the
normalized state vector with itself). That in turn implies via equation
\ref{expectation value} that the trace of ${\hat O}$ is $1$.

To sum up, one can use density operators in matrix form to represent both
states of complete and incomplete knowledge (i.e. pure or mixed states). We
saw, however, that for pure states a vector representation is sufficient. If
one wants to use the same mathematical tool to write down any state
irrespective of the knowledge he holds on its preparation procedure then the
method of choice is the density operator (also called density matrix). A system
is in a {\em pure} state when the corresponding density operator in equation
\ref{density operator} contains only one term. In this ideal case, there is no
lack of knowledge on the preparation of the system, the preparation procedure
generates the desired output with unit probability. This implies that the
diagonalized density matrix representing a pure state has all entries equal to
zero except one entry equal to $1$ on the principal diagonal. Therefore, if you
take the trace of the diagonalized density matrix squared, you will still get
one. Furthermore, the trace of the diagonalized density matrix squared is equal
to the trace of the original density matrix squared (remember the trace is
invariant under unitary transformations). This observation is the basis of a
criterion to check whether a given density matrix represents a pure or a mixed
state. The test consists in taking the trace of the density matrix squared. If
the trace is equal to $1$, then the state is pure otherwise it is mixed. We
recall that a {\em mixed} state arises in situation when the preparation
procedure is faulty and the result is a distribution of different outputs each
occurring with a given probability.

\subsubsection{Mixed states for two quantum systems}

Our treatment of density operators for single quantum systems can be applied to
bipartite systems with no essential modification. Let us consider an example in
which an experimental apparatus produces the maximally entangled state
$|\psi_{AB}\rangle$ (see equation \ref{maxent}) with probability $p_0$ and the
product state $\ket{0}_A\ket{0}_B$ with probability $p_1$. For both states we
know how to construct the corresponding projectors by using the same method
illustrated in equation \ref{projent}. But, before writing down the resulting
density operator, we introduce a small simplification in the notation used. We
write the state $\ket{0}_A\ket{0}_B$ simply as $\ket{00}_{AB}$ or simply
$\ket{00}$. The rule to write down the four-dimensional vector corresponding to
this state and its interpretation does not change. The first digit still refers
to atom $A$ and the second to atom $B$. We can now write the corresponding
density operator ${\hat \rho_{AB}}$ as shown in equation \ref{density operator}

\begin{eqnarray}
{\hat \rho_{AB}}&=& p_0 \projector{\psi_{AB}} +  p_1 \projector{00}\;\;
    \nonumber\\
    &=&
    \frac{p_0}{2}\left( \begin{array}{c c c c}
     0 & 0 & 0 & 0 \\
         0 & 1 & 1 & 0 \\
         0 & 1 & 1 & 0 \\
         0 & 0 & 0 & 0
    \end{array} \right) + p_1 \left( \begin{array}{c c c c}
           1 & 0 & 0 & 0 \\
         0 & 0 & 0 & 0 \\
         0 & 0 & 0 & 0 \\
         0 & 0 & 0 & 0
    \end{array} \right) \;\;
       \nonumber\\
    &=&
    \left( \begin{array}{c c c c}
     p_1 & 0 & 0 & 0 \\
         0 & \frac{p_0}{2} & \frac{p_0}{2} & 0 \\
         0 & \frac{p_0}{2} & \frac{p_0}{2} & 0 \\
         0 & 0 & 0 & 0
    \end{array} \right) \; .
\label{mix1}
\end{eqnarray}
There is another situation that will arise in later sections. Suppose that two
distant machines are generating one atom each, but we do not know exactly the
preparation procedure of each atom. Since the two machines are very far away
from each other, we can ignore the interaction between the atoms and describe
them separately in two different 2-dimensional Hilbert spaces by writing down
the corresponding single particle density operators ${\hat \rho_A}$ and ${\hat
\rho_B}$. All this is fine. But, we may also write the joint state of these two
non-interacting atoms as a density operator ${\hat \rho_{AB}}$ in our
4-dimensional Hilbert space, as we did for the case considered in equation
\ref{mix1}. How do we proceed? We simply take the tensor product between the
two $2 \times 2$ matrices corresponding to ${\hat \rho_A}$ and ${\hat \rho_B}$
to get

\begin{equation}
{\hat \rho_{AB}}= {\hat \rho_A} \otimes {\hat \rho_B}
\label{mix2}
\end{equation}
We leave as an exercise for the reader to choose two arbitrary density
operators ${\hat \rho_A}$ and ${\hat \rho_B}$ and perform an explicit
calculation of ${\hat \rho_{AB}}$.

Once we know how to write 1) the density matrix for the joint
state of the two atoms and 2) the matrix representing a joint
observable or projector we will have no trouble finding
expectation values or probabilities of certain measurement
outcomes. All we need to do is to multiply two $4\times 4$
matrices and take the trace as illustrated for a single particle
in equations \ref{expectation value} and \ref{prob proj2}.

\subsubsection{The reduced density operator}
\label{Thereduced}

There is another context in which a mixed state arises even when there is no
uncertainty in the preparation procedure of the quantum system one is holding.
Imagine you have an ideal machine that generates, with probability one, {\em
pairs} of maximally entangled particles in the state
$\ket{\psi_{AB}}=\frac{1}{\sqrt{2}}(|00\rangle+|11\rangle)$. The density
operator $\rho_{AB}$ for this pure state reduces to the corresponding
projector, because all the probabilities except one are vanishing see
discussion at the end of section \ref{Thecreation}. In fact, the $4 \times 4$
density matrix for this preparation procedure was explicitly calculated in
equation \ref{projent}.

After having created the entangled pair we decide to lock particle
$A$ in a room to which we have no access and we give particle $B$
to our friend Bob. Bob can do any measurement he wants on particle
$B$ and he would like to be able to predict the outcomes of any of
these. Evidently Bob does not know what is happening to particle A
after it has been locked away and as a consequence {\em now} he
has an incomplete knowledge of the total state. The question is
how we can describe mathematically his state given the incomplete
knowledge that Bob has of particle $A$. The first point to make is
that Bob still has some background knowledge on particle $A$
because he retains information on the original preparation
procedure of the entangle pair. For example, he knows that if
Alice subjects her particle to an energy measurement and finds
that particle $A$ is in the ground (excited) state, then particle
B has to be in the ground (excited) state too. This prediction is
possible because the measurement outcomes of the two particles are
always correlated because they were prepared in the entangled
state $\ket{\psi_{AB}}$. Furthermore, Bob knows from the
preparation procedure, that the probability that Alice finds her
particle in either the ground state $\ket{0}_A$ or in the excited
state $\ket{1}_A$ is $\frac{1}{2}$. By using the non local
correlations between his particle and the other, Bob concludes
that particle $B$ too is in either the ground state $\ket{0}_B$ or
in the excited state $\ket{1}_B$ with probability $\frac{1}{2}$.
Now let us assume that Alice indeed has measured the energy
operator on her particle but, as she is inside the box, has not
told Bob that she did this. Therefore, in half the cases Bob's
particle will be in state $|0\rangle\langle 0|$ and in half the
cases it will be in state $|1\rangle\langle 1|$. This is a
situation that is most easily described by a density operator. We
find that the state of Bob's particle is described by the reduced
density operator ${\hat \rho_B}$ given by:

\begin{eqnarray}
    {\hat \rho_B} &=& \frac{1}{2}\projector{0} + \frac{1}{2}\projector{1} \;\;.
    \nonumber\\
    &=&
    \frac{1}{2}\left( \begin{array}{c c}
    1 & 0 \\
    0 & 1
    \end{array} \right) \;\; .
    \label{redope}
\end{eqnarray}
where we used the rules for the representation and manipulation of quantum
states as vectors (equation \ref{Mabuchi5.3}). From the above reasoning it is
perhaps not surprising that ${\hat \rho_B}$ is often termed the reduced density
operator. Being a mixed state, it represents Bob's incomplete information on
the state of his particle (the reduced system) due to his inability to access
particle A while the total system is in a pure entangled state represented by
the larger matrix ${\hat \rho_{AB}}$. In fact, Bob wrote down ${\hat \rho_B}$
after taking into account all information that was available to him. It is
important to note that we would have obtained the same result for Bob's density
operator if we had assumed any other operation on Alice's side. The key point
is that, as Alice's actions do not affect Bob's particle in any physically
detectable way, it should not make any difference for Bob's description of his
state which assumptions he makes for Alice's action.

The whole operation of ignoring Alice's part of the system and
generating a reduced density operator only for Bob's system is
sometimes written mathematically as

\begin{equation}
    {\hat \rho_A}=tr_B\{{\hat \rho_{AB}}\} \; .
    \label{partial trace}
\end{equation}
The mathematical operations that one has to perform on the entries of the
larger matrix ${\hat \rho_{AB}}$ in order to obtain ${\hat \rho_A}$ are called
the partial trace over system B. The general case can be dealt with analogously
to the reasoning above. One assumes that in the inaccessible system a
measurement is carried out whose outcomes are not revealed to us. We then
determine the state of our system for any specific outcome from the projection
postulate and we use the associated probabilities to form the appropriate
density operator. We refer the reader interested in learning how to deal with
this method in the most efficient way to some recent courses of quantum
mechanics \cite{Plenio notes,Mabuchi notes,Preskill notes}.

This topic concludes our very concise review of quantum mechanics. We will now
extensively apply the mathematical tools introduced in this section to deal
with situations in which classical information is encoded in a quantum system
and later to discuss the new field of quantum information theory. It is
therefore essential that the reader feels confident with what he has learned so
far before moving on.

\section{Classical information encoded in quantum systems}
\label{Classicalinformationencodedinquantumsystems}

\subsection{How many bits can we encode in a quantum state?}
\label{Howmanybits}

In the previous section, we studied two situations in which the state of a
quantum system is mixed, namely when the preparation procedure is not
completely known or when we have a subsystem that is part of a larger
inaccessible system. In both cases, our knowledge was limited to the
probabilities $\{p_i\}$ that the system is in one of the pure states
$\ket{\psi_i}$. A question that arises naturally in this context is whether we
can assign an entropy to a quantum system in a mixed state in very much the
same way as we do with a classical system that can be in a number of
distinguishable configurations with a given set of probabilities. In the
classical case the answer is the well known Boltzmann formula given in equation
\ref{Boltzmann entropy}. At first sight, you may think that the same formula
can be applied to evaluate the mixed state entropy just by plugging in the
probabilities $\{p_i\}$ that the quantum system is in one of the pure quantum
states $\ket{\psi_i}$. Unfortunately, this idea does not work, because the
quantum states $\ket{\psi_i}$ are different from the distinguishable
configuration of a classical system in one important way. They are not always
perfectly distinguishable! As we pointed out earlier, two quantum states can be
non-orthogonal and therefore not perfectly distinguishable. But maybe the idea
of starting from the classical case as a guide to solve our quantum problem is
not that bad after all.

In particular, imagine that you are given the density matrix representing the
mixed state of a quantum system. Can you perform some mathematical operations
on this matrix to bring it in a form that is more suggestive? You may recall
from equations \ref{Mabuchi5.3} and \ref{density operator} that the procedure
to write down this density matrix is the following. First construct the matrix
representation of the projector $\projector{\psi_i}$ for each of the vectors
$\ket{\psi_i}$, then multiply each of them by their respective probability and
finally sum all up in one matrix. The reader can check that the prescription on
how to construct each matrix $\projector{\psi_i}$ given in equation
\ref{Mabuchi5.3} ensures that the resulting density matrix is Hermitian. We
denote the orthogonal eigenvectors of our (hermitian) density matrix by
$\ket{e_i}$. If we choose the $\ket{e_i}$ as basis vectors, we can rewrite our
matrix in a diagonal form. All the entries on the diagonal are the real
eigenvalues of the matrix. These matrices can now be written in Dirac notation as

\begin{equation}
    {\hat\rho} = \sum q_i |e_i\rangle\langle e_i| \;\; ,
\label{diagonal}
\end{equation}
where the $q_i$ are now the eigenvalues of the density matrix. This new matrix
actually represents another preparation procedure namely the mixed state of a
quantum system which can be in any of the $orthogonal$ states $\ket{e_i}$ with
probability $q_i$. But now the states $\ket{e_i}$ are distinguishable and
therefore one can apply the Boltzmann formula by simply plugging the
eigenvalues of the matrix as the probabilities.

There is one problem in this reasoning. When you rewrite the old density matrix
in diagonal form you are actually writing down a different matrix and therefore
a representation of a different preparation procedure. How can you expect then
that the entropy so found applies to the mixed state you considered originally?
The answer to this question lies in the fact that what matters in the matrix
representation of quantum mechanical observables or states is not the actual
matrix itself, but only those properties of the matrix that are directly
connected to what you can observe in the lab. From the previous section, we
know that all the physically relevant properties are basis independent. The
diagonalization procedure mentioned above is nothing else than a change of
basis and therefore there is no harm in reducing our original density matrix
$\hat{\rho}$ in diagonal form and hence define the von Neumann entropy as the
function

\begin{eqnarray}
    S({\hat\rho}) &=& -tr\{{\hat\rho}log{\hat\rho}\} \;\;
    \nonumber\\
    &=& - \sum q_i \log q_i \;\; . \label{Von Neumann}
\end{eqnarray}
The formula above is an example of how a function of a matrix can be evaluated
as an ordinary function of its eigenvalues only. Since the eigenvalues are
invariant under a change of basis the function itself is invariant, as
expected. One can check the validity of the formula above as an entropy measure
by considering two limiting cases. Consider first a pure state, for which there
is no uncertainty on the output of the preparation procedure. The probability
distribution reduces to only one probability which is one. Therefore the
density matrix representing this state has eigenvalue equal to one. If you plug
the number one in the logarithm in formula \ref{Von Neumann} you get the
reassuring result that the entropy of this state is zero. On the other hand,
for a maximally mixed state in which the system can be prepared randomly in one
of $N$ equally likely pure state we find that the entropy is $log N$ in
agreement (in dimensionless units) with the Boltzmann and Shannon entropies.

There is an interesting point to note. If we create a mixed state by generating
the states $\{|\psi_i\rangle\}$ with probabilities $\{p_i\}$ we first hold a
list of numbers which tell us which system is in which quantum state. In this
classical list each letter holds $H(\{p_i\})$ bits of information. If we want
to complete the creation of the mixed states, we have to erase this list and,
according to Landauer's principle, will generate $kT H(\{p_i\})$ of heat per
erased message letter. In general the Shannon entropy is larger than or equal
to the von Neumann entropy of the density operator $\hat\rho=\sum_i p_i
|\psi_i\rangle\langle\psi_i |$. It is also clear that the same mixed state can
be created in many different ways and that the information invested into the
state will not be unique. It seems therefore unclear whether we can ascribe a
unique classical information content to a mixed state. However, the only
quantity that is independent of the particular way in which the mixed state has
been generated is the von Neumann entropy which is different from the amount of
information invested in the creation of the mixed state. In fact, the von
Neumann entropy $S(\hat\rho)$ is the {\em smallest} amount of information that
needs to be invested to create the mixed state $\hat\rho$. As we are unable to
distinguish different preparations of the same density operator ${\hat \rho}$
this is certainly the minimum amount of classical information in the state
$\hat\rho$ that we can access. The question is whether we can access even more
classical information. The answer to this question is $NO$, as we will see in
the next section in which we generalize Landauers principle to the quantum
domain to illuminate the situation further. The result of these considerations
is that there is a difference between information that went into a mixed state,
and the {\em accessible} information that is left after the preparation of the
states \cite{Peres 95}.

\subsection{Erasing classical information from quantum states: Landauer's principle revisited}
\label{Erasingclassicalinformationfrom quantum states}

In the previous subsection we have discussed the amount of classical
information that goes into the creation of a mixed state. But an obvious
question has not been discussed yet: {\it how do you erase the classical
information encoded in a quantum mixed state ?} In section
\ref{Erasingclassicalinformationfromclassicalsystems}, we explained how to
erase one bit encoded in a partitioned box filled with a one molecule gas. All
you have to do in this simple case is to remove the partition and compress the
gas on one side of the box (say the right) independently of where it was
before. This procedure erases the classical state of the binary device and the
bit of information encoded in it. If the compression is carried out reversibly
and at constant temperature, then the total change of thermodynamical entropy
is given by $kln2$, the minimum amount allowed by Landauer's principle. In this
sense the erasure is optimal. What we are looking for in this section is a
procedures for the erasure of the state of {\em quantum systems}. We will first
present a direct generalization of the classical erasure procedure and then
follow this up with a more general procedure that applies directly to both
classical and quantum systems. These results will then be used to show that the
accessible information in a quantum state $\hat\rho$ created from an ensemble
of pure states is equal to $S(\hat\rho)$.

\subsubsection{Erasure involving measurement}

We know from the previous section that the information content of a pure state
is zero. Therefore, all we need to do to erase the information encoded in a
mixed quantum state, is to return the system to a fixed pure state called the
standard state. We show how to achieve this in the context of an example.

\begin{figure}[hbt]
\begin{center}
\epsfxsize9.0cm \centerline{\epsfbox{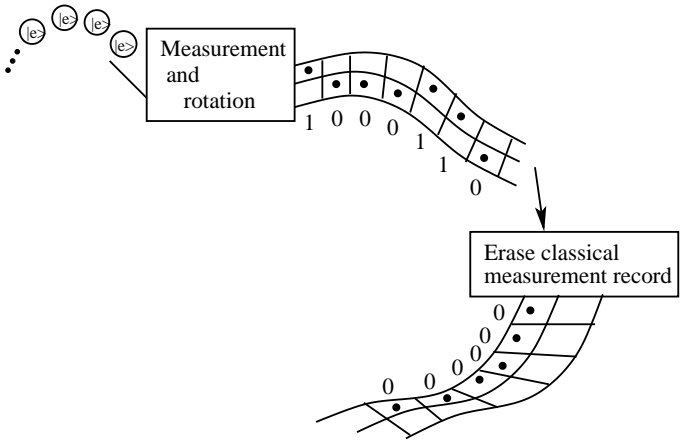} }
\end{center}
\vspace*{0.15cm}
 \caption{\label{quantumerase1} Particles described by a quantum state $\rho$
arrive and are being measured in a basis $|e_i\rangle$ giving the
outcome $i$ with probability $p_i$. Given the outcome the each of
the particles can be rotated into the pure state $|e_0\rangle$.
The remaining classical list has to be erased as well. This
generates $kT ln 2 H(\{p_i\})$ of heat. This procedure can be
optimized if one measures in the eigenbasis of $\rho$ in which
case one generates $kT ln2 S(\rho)$ heat.}
\end{figure}

Imagine you want to erase the information encoded in quantum systems in the
mixed state $\rho=\sum_{i} p_i |e_i\rangle\langle e_i|$ where the $\ket{e_i}$
are the energy eigenstates. You start by performing measurements in the energy
eigenbasis . After the measurement is performed, each system will indeed be in
one of the pure states $\ket{e_i}$ and we have a classical record describing
the measurement outcomes. If the density operator represents the preparation
procedure of two level atoms and we measure their energies, the classical
measurement record would be a set of partitioned boxes storing a list of 0s and
1s labeling the energy of the ground state or the excited state for each atom
measured. Now we can apply a unitary transformation and map the state
$|e_i\rangle$ onto the standard state $|e_0\rangle$ for each atom on which a
measurement has been performed (see first step in figure \ref{quantumerase1}).
Naively, one could think that this completes the erasure, because we have reset
the quantum systems to a fixed standard state $|e_0\rangle$. However this is
not true, because we are still holding the classical measurement records so the
erasure is still not complete. We need one more step namely to erase the
classical measurement record using the classical procedure discussed above. In
the example of figure \ref{quantumerase1} , this amounts to compressing each of
the partitioned boxes where the list of 0s and 1s were encoded. This process
will generate an amount of thermodynamical entropy not less than $kln2$ per
bit. In general we have that $k \ln 2 S(\hat\rho) \le k \ln 2 H(p)$ as pointed
out in the previous section. The optimal erasure procedure, ie the one that
creates the least amount of heat, is the one where the quantum measurements are
made in the basis of the eigenstates of $\hat\rho$, so that the Shannon entropy
equals the von Neumann entropy as discussed in section \ref{Howmanybits} .

To sum up, the protocol described above relies on a quantum
mechanical measurement followed by a unitary transformation and
the erasure of the classical measurement record. While this
protocol is a perfectly acceptable erasure procedure, it consists
of two conceptually different steps and one may wonder whether
there is a simpler method that does not involve the explicit act
of measuring the quantum system.

\subsubsection{Erasure by thermal randomization}

Such an elegant way to erase information, which has been introduced by Lubkin
\cite{Lubkin 87,Vedral 00}, is by thermal randomization. Simply stated, you
have to place the quantum system that is to be erased into contact with a heat
bath at temperature $T$. The laws of statistical mechanics teach us that when
thermal equilibrium is reached, there will be an uncertainty about the energy
state the system is in. The origin of this uncertainty is classical because it
is induced by thermal fluctuations. This situation of lack of knowledge of the
preparation procedure for the quantum state is equivalent to the example of the
oven considered in section \ref{Thecreation}. The state of the system can
therefore be written as a density operator ${\hat \omega}$ given by

\begin{eqnarray}
    {\hat \omega} &=& \frac{e^{-\beta {\hat H}}}{Z} \nonumber\\
    &=& \frac{\sum_{i} e^{-\beta {E_i}} |e_i\rangle\langle
    e_i|}{Z} \;\; ,
\label{Boltzmann}
\end{eqnarray}
where $\beta=1/kT$, ${\hat H}$ is the Hamiltonian of the system whose
eigenstates and eigenvalues are $\ket{e_i}$ and $E_i$ respectively. The number
$Z$ is the partition function of the system and can be calculated from
$Z=tr\{e^{-\beta {\hat H}}\}$. For example, the system can be in its ground
state with probability $p_0$ given by the Boltzmann distribution:

\begin{equation}
    \ p_0 = \frac{e^{-\beta E_0}}{Z} \;\; .
\end{equation}
The exponential dependence of the probabilities in the equation above implies
that, if the system has a sufficiently large level spacing (ie $E_0$ is much
smaller than the other energy levels), it will be almost surely in its ground
state. Thus, if a measurement is made, the result will be almost certainly that
the apparatus is in its ground state. In other words, the mixed state
${\hat\rho}$ can be made arbitrarily close to a standard pure state $\ket{e_0}$
by greatly reducing the presence of the other pure states $\ket{e_i}$ in the
thermal preparation procedure. In practice, this is exactly what we wanted: a
procedure that always resets our system, originally in the mixed state
${\hat\rho}$, to a standard state (independent of the initial state), eg the
ground state $\ket{e_0}$. Also note that this erasure procedure never requires
any measurement to be performed, so we do not need to be concerned with erasing
the classical measurement record, as in the previous method.

Furthermore, we can readily calculate the {\em net} amount of
thermodynamical entropy generated in erasing the quantum mixed
state  ${\hat \rho}$ where the classical information is encoded.
We proceed by computing first the change of thermodynamical
entropy in the system and then the change of thermodynamical
entropy of the environment. All the steps in this derivation are
reproduced and motivated. The readers who do not feel comfortable
with the formalism of density operators explained in the previous
sections can skip this derivation and jump to the result in
equation \ref{quantum eraser}.

\begin{figure}[hbt]
\begin{center}
\epsfxsize8.0cm \centerline{\epsfbox{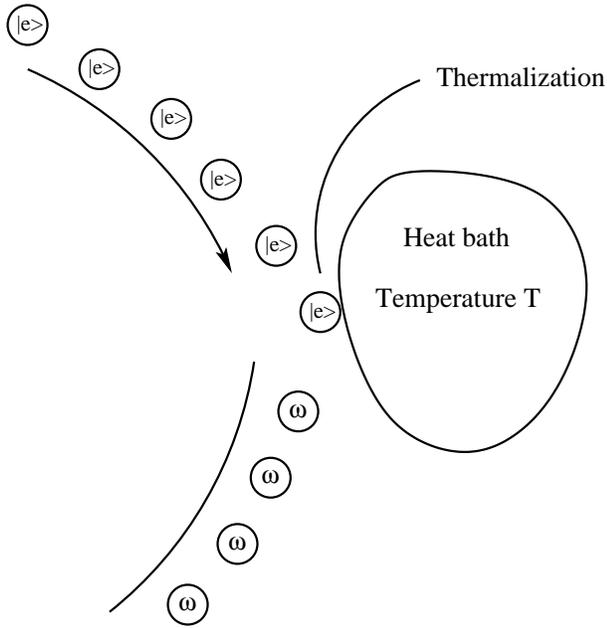} }
\end{center}
\caption{\label{quantumerase2} The quantum particles, described by
the average state $\rho$, are brought into contact with a thermal
heat bath and are allowed to relax into thermal equilibrium. The
resulting change of heat depends on the temperature of the heat
bath and its optimal value is given by $kT ln 2 S(\rho)$ }
\end{figure}

The mixed state $\hat \rho$ is generated by a source that produces randomly
pure states $|e_i\rangle$ with probability $p_i$. {\em Each} quantum system in
such a pure state $|e_i\rangle$ is brought into contact with the heat bath and
thermalizes into the state $\hat \omega$ (see figure \ref{quantumerase2}).

We remind the reader that the entropy of the system before the thermalization
procedure takes place is zero because the system is in one of the pure states
$|e_i\rangle$ (see equation \ref{Von Neumann} and discussion below). Therefore,
in each of these contacts, the thermodynamical entropy of the system increases
by the same amount $kln2 S(\hat\omega)$, where $S(\hat\omega)$ is the von
Neumann entropy times the conversion factor between information and
thermodynamical entropy, so that we have

\begin{equation}
    \Delta S_{sys} = kln2 S(\hat\omega)\; .
\end{equation}
Now we proceed to discuss the change in the thermodynamical entropy of the heat
bath. The latter is given in terms of the heat lost by the heat bath and its
temperature $T$ by the well known thermodynamical relation $\Delta
S_{bath}=\frac{\Delta Q_{bath}}{T}$. The easiest way to attack this problem is
by using the observation that the change of heat in the heat bath $\Delta
Q_{bath}$ is equal and opposite to the change of heat in the system $\Delta
Q_{system}$. The latter is given in terms of the heat lost by the system and
the temperature of the reservoir by the well known thermodynamical relation $T
\Delta S_{system}=\Delta Q_{system}$. Furthermore, the first law of
thermodynamics can be used to write $\Delta Q_{system}$ as the change in the
internal energy of the system $\Delta U_{system} = U_{final}-U_{initial}$ (i.e.
the procedure can be done reversibly so that the work required is arbitrary
close to $0$). One can summarize what is stated above in the equation:

\begin{equation}
    \Delta S_{bath} = - \frac{\Delta U_{system}}{T}= -\frac{U_{final}-U_{initial}}{T} \; .
\label{calculation1}
\end{equation}
We can now rewrite the initial and final energy of the {\em system} as the
expectation value of the Hamiltonian ${\hat H}$ of the system calculated in the
initial state ${\hat \rho}$ and in the final thermal state ${\hat \omega}$. The
formula to use is given in equation \ref{expectation value}. Once this is done
equation \ref{calculation1} can be recast in the following form:

\begin{eqnarray}
    \Delta S_{bath} &=& -\frac{ tr\{ {\hat \omega}{\hat H} \} - tr\{ {\hat \rho}{\hat H}\} }{T}
    \;\; \nonumber \\
    &=& -\frac{ tr\{ ({\hat \omega}-{\hat \rho}) {\hat H} \} }{T}\;\;.
    \label{calculation2}
\end{eqnarray}
The expression in equation \ref{calculation2} can be further elaborated by
substituting the operator ${\hat H}$ with the corresponding expression $-kT
ln(Z{\hat \omega})$ obtained after solving the first equation in
\ref{Boltzmann} with respect to ${\hat H}$.

\begin{eqnarray}
    \Delta S_{bath} &=& k tr\{({\hat \omega}-{\hat \rho})ln(Z{\hat \omega})\}\;\;
\nonumber\\
    &=&  k tr\{({\hat \omega}-{\hat \rho})ln{\hat \omega}\}+ k lnZ tr\{({\hat \omega}-{\hat \rho})
ln{\hat \omega}\}\;\;. \label{calculation2a}
\end{eqnarray}
In the previous steps we used the properties of logarithm and the fact that a
constant like $lnZ$ or $kT$ can be "taken out of the trace". The last term in
equation \ref{calculation2a} vanishes because $tr\{\rho\}=tr\{\omega\}=1$
because the trace of a density operator is always equal to $1$. Also the first
term can be expanded as

\begin{eqnarray}
    \Delta S_{bath} &=& k tr\{ {\hat \omega}ln{\hat \omega}\} - k tr\{ {\hat \rho}ln{\hat \omega} \}\;\;
\nonumber\\
    &=& - kln2 S({\hat \omega}) - k tr\{ {\hat \rho} ln {\hat \omega } \}\;\;.
\label{calculation3}
\end{eqnarray}
Note the factor $ln2$ to convert the logarithm from the natural basis to the
basis $2$ adopted in the definition of the Von Neumann entropy. We therefore
reach the final result that the total change of thermodynamical entropy in
system and environment in our procedure is given by

\begin{equation}
    \Delta S_{tot} = \Delta S_{sys} + \Delta S_{bath} = - k tr\{ {\hat \rho}ln {\hat \omega}
    \},
\label{quantum eraser}
\end{equation}
where ${\hat \omega}$ is the state of the system after having reached thermal
equilibrium with a heat bath at temperature $T$.

This entropy of erasure can be minimized by choosing the temperature of the
heat bath such that the thermal equilibrium state of the system is $\hat\rho$,
i.e.
\begin{equation}
    min\{\Delta S_{tot}\} = S({\hat \rho}) = - tr\{{\hat \rho}log{\hat \rho}\}
    \; ,
\label{minimum}
\end{equation}
which equals the von Neumann entropy of $\hat\rho$. Equation \ref{minimum}
restates Landauer's principle for quantum systems in which classical
information is encoded.

From the last section we remember that the amount of classical information
invested in the creation of the state $\hat\rho$ was never smaller than the von
Neumann entropy $S(\hat\rho)$ a value which can always be achieved. This left
open the question how much classical information is actually still accessible
after the creation of $\hat\rho$. Having seen above, that the entropy of
erasure of a quantum state $\hat\rho$ can be as small as the von Neumann
entropy we conclude from Landauer's principle, that the accessible information
in the state $\hat\rho$ cannot be larger than its von Neumann entropy.
Therefore it becomes clear that the only possible quantity to describe the
classical information content of a mixed state that has been prepared from an
ensemble of pure states is given by the von Neumann entropy.

\subsection{Classical information transmitted through a noisy quantum channel}
\label{Classicalinformationtransmittedthroughanoisyquantumchannel}

In this section we will evaluate how much classical information can be
transmitted reliably down a {\em noisy} quantum channel. The reader may
remember that we considered the classical analogue of this problem in section
\ref{Sendingclassicalinformationthroughnoisyclassicalchannel}.

Imagine that Alice wants to transmit a message to Bob. This
message is written in an alphabet composed of $N$ letters $a_i$
each occurring with probability $p_i$. Alice decides to encode
each letter $a_i$ in the pure quantum state $\ket{\psi_i}$. Alice
can transmit the letter $a_i$ simply by sending a particle in the
state $\ket{\psi_i}$ via a physical channel, like an optical
fiber. When Bob receives the particle, he does not know which pure
state it is in. Bob's incomplete knowledge of the state of the
particle is represented by the mixed state ${\hat \rho}= \sum p_i
\projector{\psi_i}$. When Bob reads the state of the particle he
will have gained some useful information to guess which letter
Alice had encoded. The information encoded in the mixed state of
the quantum carrier is equal to the von Neumann entropy $S(\rho)$
as explained in the last section. If the states $\ket{\psi_i}$ are
orthogonal, then the von Neumann entropy reduces to the Shannon
entropy of the probability distribution $\{ p_i\}$  because all
the quantum states are distinguishable and the situation is
analogous to the classical case. If the states are non-orthogonal
then the von Neumann entropy will be less that the Shannon
entropy. The information transfer is degraded by the lack of
complete distinguishability between the {\em pure} states of the
carriers in which the information was encoded at the source. This
feature has no classical analogue and is sometimes referred as
intrinsic quantum noise. The name is also justified by the fact
that this $noise$ is not induced by the environment or any
classical uncertainty about the preparation procedure of the
carriers' states.

\begin{figure}[hbt]
\begin{center}
\epsfxsize8.0cm \centerline{\epsfbox{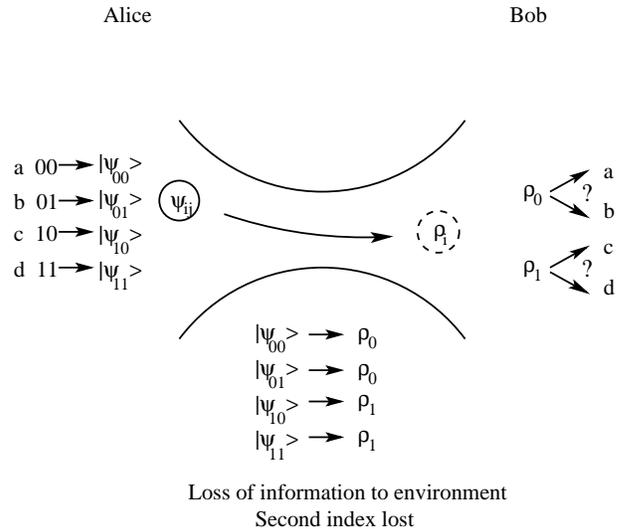} }
\end{center}
\caption{\label{channel} The basics of information transmission.
Alice encodes the letters a,b,c,d (which can also be encoded in
binary as $00,01,10,11$) and encodes them in pure quantum states
$\psi_{ij}\rangle$. These states are sent through the channel
where the environment interacts with them. Here the information
about the second index is lost leading to mixed states $\rho_0$
and $\rho_1$. Bob receives these mixed states and has lost some of
the original information as he cannot distinguish between a and b
and between c and d. }
\end{figure}

We now wonder what happens when the channel itself is noisy (see
figure \ref{channel}). For example, the optical fiber where the
carriers travel could be in an environment or an eavesdropper,
Eve, could be interacting with the carriers. This extra noise is
not intrinsic to the preparation of the pure states at the source,
but it is induced by the environment. One can view the
transmission through a noisy channel in the following way.
Initially the sender, Alice, holds a long classical message. She
encodes letter $i$ (which appears with probability $p_i$) of this
message into a $pure$ state that, during the transmission, is
turned into a possibly mixed quantum state $\rho_i$ due to the
incomplete knowledge of the environment or of Eve's actions. These
quantum states are then passed on to the receiver, Bob, who then
has the task to infer Alice's classical message from these quantum
states. The upper bound for the capacity for such a transmission,
i.e. the information $I$ that Bob can obtain about Alice's message
per sent quantum state, is known as the Holevo bound

\begin{equation}
    I = I_H = S(\rho) - \sum_i p_i S(\rho_i) \;\; ,
    \label{Holevo}
\end{equation}

The rigorous proof of this result is rather complicated
complicated. \cite{Holevo 98}. The aim of the next section is to
$justify$ Holevo's bound from the assumption of the validity of
Landauer's principle.

\subsubsection{Holevo's bound from Landauer's principle}
\label{Holevobound}

The idea behind the derivation of the Holevo bound from Landauer's
principle is to determine an upper bound on the entropy that is
generated when Bob erases the information that the message system
carries in its state $\rho_i$. In this way we directly obtain an
upper bound on the information received by Bob, because we know
from Landauer's principle that the information received is always
less or equal to the entropy generated when it is erased (see
equation \ref{minimum}).

Let us begin by assuming that Alice uses an alphabet of letters
$(i,\alpha)$ that are enumerated by the two integers $i$ and
$\alpha$. We use this form of double indices to make formulation
of the following analysis simpler, but apart from that it has no
deeper meaning. The letter $i$ appears with probability $p_i$ and
given $i$, $\alpha$ appears with the probability $r_{\alpha}^i$.
Alice encodes her message in the following way. Given she wants to
send letter $(i,\alpha)$ which occurs with probability $p_i\cdot
r_{\alpha}^i$, she encodes it into the pure state
$|\phi_{\alpha}^{i}\rangle$. Therefore $\rho_i= \sum_{\alpha}
r_{\alpha}^i |\phi_{\alpha}^{i}\rangle\langle\phi_{\alpha}^{i}|$.
Now these quantum states are inserted into the quantum channel and
they are subjected to an interaction with the environment or an
eavesdropper Eve. The effect of this interaction is that the
systems loose their correlation to the specific values of $\alpha$
or in other words, the information about $\alpha$ is lost, and we
are left with a certain degree of correlation between the integers
$i$ and the mixed states $\rho_i$. Evidently the lost information
about $\alpha$ has leaked into the environment or to Eve and this
information is not available to Bob anymore. In the following we
would like to compute, using Landauers principle, how much
information has actually been lost. To this end we construct an
optimal erasure procedure and compute the thermodynamical heat it
generates.

\subsubsection{Direct erasure}
\label{Directerasure}

As explained above message letter $(i,\alpha)$ which appears with
probability $p_i\cdot r_{\alpha}^i$ is encoded in state
$|\phi_{\alpha}^i\rangle$. We will now delete the information
encoded in these pure state by bringing them into contact with a
heat bath. We chose the temperature of this heat bath such that
the thermal equilibrium state of the message system is
$\rho=\sum_i p_i \rho_i$. This ensures that the erasure is
optimal, in the sense that it produces the smallest possible
amount of heat. Following the analysis of Lubkin's erasure in
section \ref{Erasingclassicalinformationfrom quantum states}, the
entropy of erasure is given by
\begin{equation}
    \Delta S_{er}^{(2)} = -\sum_i p_i tr\{\rho_i\log\rho\} = S(\rho) \;\; .
\end{equation}
Note that all information has been deleted because now every
quantum system is in the same state $\rho$ so that there is no
correlation between the original letter $i$ and the encoded
quantum state left after the erasure!

\subsubsection{Two step erasure}
\label{Twosteperasure}

Now let us compute the entropy of erasure in going from the pure
states $|\phi^i_{\alpha}\rangle$ into which Alice encoded her
message initially to the mixed states $\rho_i$ that Bob obtains
after the carriers have passed the channel. This is the first step
in our erasure procedure and determines the amount of information
lost to the environment or the eavesdropper.

For a fixed $i$ which appears with probability $p_i$, we place the
encoded pure states into contact with a heat bath. The temperature
$T$ of the heat bath is chosen such that the thermal equilibrium
state of the message system is $\rho_i$. Again this choice ensures
that the erasure is optimal. According to our analysis of the
Lubkin erasure in section \ref{Erasingclassicalinformationfrom
quantum states}, the entropy of erasure is then found to be
\begin{eqnarray}
    \Delta S_{er}^{(1)} &=& -\sum_i p_i \sum_{\alpha}
    tr\{ r_{\alpha}^i |\phi_{\alpha}^i\rangle\langle\phi_{\alpha}^i| \log\rho_i\} \nonumber\\
    &=& -\sum_i p_i tr\{\rho_i\log\rho_i\} \nonumber\\
      &=& \sum_i p_i S(\rho_i) \;\; .
\end{eqnarray}
After this first step in the erasure procedure there is still some
information left in the physical systems as the letter $i$ of the
classical message is correlated with the state $\rho_i$ of the
quantum system. Therefore some information is available to Bob. In
fact, this is exactly the situation in which Bob is after he
received a message which is encoded as in mixed states $\rho_i$.
To obtain a bound on the information that Bob is now holding, we
need to find a bound on the entropy of erasure of his quantum
systems.

Now we would like to determine the entropy of erasure of the
signal states $\rho_i$ that Bob has received through the channel.
In order to carry out this second step of the erasure procedure we
place each of Bob's systems, which is in one of the states
$\rho_i$ with probability $p_i$, into contact with a heat bath
such that the thermal equilibrium state of the message system is
$\rho$. As the average state of the systems is $\rho=\sum_i p_i
\rho_i$, we expect the erasure to be optimal again. We can see
easily  that this second step of erasure, just generates an amount
of entropy that is the difference between the entropy of erasure
of the first procedure and that of the first step of the second
procedure. Therefore the entropy of erasure of Bob's systems which
are in one of the states $\rho_i$'s is
\begin{eqnarray}
    \Delta S_{er}(Bob) &=& \Delta S_{er}^{(2)} - \Delta S_{er}^{(1)} \nonumber \\
    &=& S(\rho) - \sum_i p_i S(\rho_i) \;\; .
\end{eqnarray}
As the largest possible amount of information available to the
receiver Bob is bounded by his entropy of erasure we have
\begin{equation}
    I \le \Delta S_{er}(Bob) = S(\rho) - \sum_i p_i S(\rho_i) = I_H \;\; .
\end{equation}
Therefore we have obtained the Holevo bound on the information in
the states $\rho_i$ which appear with probabilities $p_i$.
 The
Holevo bound completes our answer to the first of the three
questions posed in the introduction. This is the last result that
we prove in this article about classical information. We now turn
our attention to the newly developed subject of quantum
information theory.

\section{The basics of quantum information theory}
\label{Quantuminformationencodedinquantumsystems}

The concept of quantum information represents a radical departure
from what we have encountered so far. In the next few sections, we
will explore some of its properties by using Landauer's erasure
principle. But first we want to discuss why the term {\em quantum
information} has been introduced and what exactly it means.

\subsection{Quantum information: motivation of the idea}
\label{Motivation}

The choice of the bit as the fundamental unit of information is
reasonable both logically and physically. In fact, right from the
outset, our definition of information content of an object has
focused on the fact that information is always encoded in a
physical system. Classically, the simplest physical system in
which information can be encoded is a binary device like a switch
that can be either open (1) or closed (0). However, as technology
shrinks more and more, we need to abandon the macroscopic world in
favor of devices that are sufficiently small to deserve the name
of quantum hardware. To some extent, the quantum analogue of a
classical $binary$ device is a two level quantum system like a
spin-half particle. Just as the classical device, it possesses two
perfectly distinguishable states (spin-up and spin-down) and as
such it is the simplest non-trivial quantum system. However, it
differs in one important way from the classical switch. The
general state $\ket{\psi}$ of a spin-half particle can be in an
arbitrary superposition of the state $\ket{\uparrow}_z$
corresponding to the spin of the particle being oriented upwards,
say in the positive $z$ direction, and of the state
$\ket{\downarrow}_z$ corresponding to the spin oriented downwards:

\begin{equation}
    \ket{\psi}=\alpha\ket{\downarrow}_z + \beta\ket{\uparrow}_z.
    \label{qubit}
\end{equation}

where $\alpha$ and $\beta$ are two arbitrary $complex$ numbers
such that $|\alpha|^{2}+|\beta|^{2}=1$. $|\alpha|^{2}$
($|\beta|^{2}$) are the probabilities for finding the particle
spin-up or spin-down in a measurement of the spin along the $z$
direction. By analogy with the classical bit, we define a $qubit$
as the information encoded in this two-level quantum system. An
example will elucidate the motivation behind this definition.

Imagine that you are holding a {\it complex quantum system} and
you want to send instructions to a friend of yours so that he can
reconstruct the state of the object with arbitrary precision. We
have previously mentioned that, if the necessary instructions can
be transmitted in the form of $n$ classical bits, then the {\it
classical information content of the object} is $n$ bits. Sending
$n$ bits of classical information is not difficult. We just need
to send a series of $n$ switches and our friend will read a $0$
when the switch is closed and a $1$ when it is open. He will then
process this information to recreate the state of a complex
quantum object like $n$ interacting spin-$\frac{1}{2}$ particles.
All this is fine, but it entails a number of problems. Firstly the
set of instructions may be very large even if we only want to
recreate a single qubit simply because the complex amplitudes are
real numbers. More importantly though, we are somewhat
inconsistent in trying to reduce the state of a quantum system to
classical binary choices. It would be more logical to transmit the
quantum state of the composite object by sending "quantum building
blocks". For example, we could try to send our instructions
directly in the form of simple two level quantum systems (qubits)
rather than bits encoded in classical switches. The hope is that,
if we prepare the $joint$ state of these qubits appropriately, our
friend will be able to manipulate them somehow and finally
reconstruct the state of the complex quantum object. Ben
Schumacher \cite{Schumacher 95,Preskill notes} proved that this is
indeed possible and he also provided a prescription to calculate
the $minimum$ number of qubits $m$ that our friend requires to
reconstruct an arbitrary quantum state. The existence of this
procedure allows us to establish an analogy with the classical
case and say that the $quantum$ information content of the object
is $m$ qubits. In this sense, the qubit is the basic unit of
quantum information in very much the same way as the bit is the
unit of classical information. We ask the reader to be patient and
wait for later sections, namely section
\ref{Thequantuminformationcontentofaquantumsysteminqubits}, in
which we will explain in more detail Schumacher's reasoning and
expand on some of the remarks made above. The previous arguments
should anyway convince the reader that, although the ideas of
qubit and bit have a common origin, it is worth exploring the
important differences between the two.

\subsection{The qubit}
\label{Thequbit}

The key to understand the differences between quantum and
classical information is the principle of superposition. Our
discussion below will be articulated in two points. We first
assess the implications of the superposition principle for the
state of a single spin-half particle (1 qubit) and then we move
to consider the case of a quantum system composed of $n$
spin-half particles ($n$ qubits).

\subsubsection{A single qubit}
\label{Asinglebitversusasinglequbit}

The concept of superposition of states, that plays a crucial role
in the definition of the state of a spin-half particle has no
analogue in the description of a classical switch which is either
in one state or in the other, but not in both! Naively, one could
think that the probabilistic interpretation of the coefficients
$\alpha$ and $\beta$ in the superposition of states given by
equation \ref{qubit} solves all the problems. In fact, if
$|\alpha|^{2}$ and $|\beta|^{2}$ are the probabilities for finding
the particle spin-up or spin down after the spin is measured along
the z direction, then a qubit is nothing more than a statistical
bit. That is a random variable, which can be either $0$ or $1$
with given probabilities $|\alpha|^{2}$ or $|\beta|^{2}$
respectively. This conclusion is wrong!

\begin{figure}[!htb]
    \centerline{\epsfxsize=9.cm \epsffile{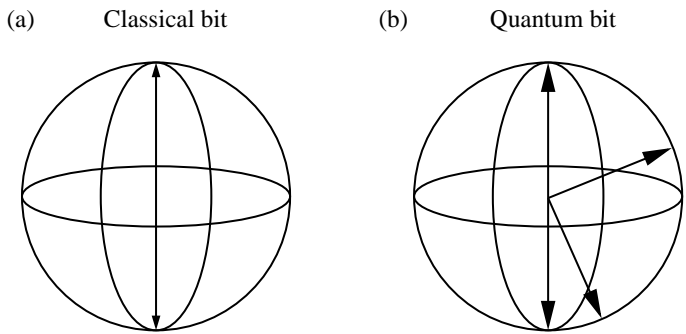}}
    \vspace*{0.15cm}
    \caption{The Bloch sphere representation of (a) a classical bit in which the
    vector can only point up or down; (b) a qubit in which the vector is allowed
    to point in any direction. This illustrates that a qubit possesses more freedom
    than a classical bit when information is $processed$.}
    \label{Blochsphere}
\end{figure}

The probabilistic interpretation of equation \ref{qubit} given
above is not the full story on the qubit because it concentrates
only on the modulus squared of the complex numbers $\alpha$ and
$\beta$. This amounts to throwing away some degrees of freedom
that are contained in the imaginary entries. We have shown before
that the qubit is mathematically described by a vector in a two
dimensional complex vector space (the Hilbert space). This state
vector can be visualized as a unit-vector in a three-dimensional
space, ie. pointing from the origin of the coordinate system to
the surface of a unit sphere, known as the Bloch sphere
\cite{Preskill notes,Scully Z 99} (see figure \ref{Blochsphere}b).

This
can be contrasted with a classical bit which is simply a discrete
variable that can take up either of the values 0 or 1. A classical
bit is thus shown in the same diagram as a unitary vector along
the $z$ axis, pointing either up or down (see figure
\ref{Blochsphere}a). This makes intuitive the idea that to some
extent there is "more room for information" in a qubit than in a
bit. However, the ability of the qubit to store more information
in its "larger space" is limited to the processing of information.
It is in fact impossible to fully access this information (ie. the
whole of the spherical surface) in a measurement. More explicitly,
whenever we manipulate a spin-up particle we do act on all its
degrees of freedom (ie. we change both the amplitude and the
relative phase of the two complex coefficients $\alpha$ and
$\beta$) so that the vector representing the qubit can be rotated
freely on any point on the surface of the sphere. However, when we
try to measure the state of the system we have to choose a basis
(ie. a direction) in which the spin measurement has to be done.
That amounts to fixing a direction in space and asking {\em only}
whether the {\it projection of the vector state} in that direction
is oriented parallel or anti-parallel. In other words when we try
to extract information from the spin-half particle we never
recover a full qubit (ie. the quantum state of the system). We
know from section \ref{nonorthogonal} that it is impossible to
extract the complex coefficients $\alpha$ and $\beta$ with a
single measurement. In fact, the information one can extract from
the measurement is just one classical bit. It is remarkable to
note, that there is a large fraction of information in a qubit
that can be processed but not accessed in a measurement.
Therefore, the difference between a single qubit and a classical
bit is not merely $quantitative$, as figure $\ref{Blochsphere}$
suggests, but also $qualitative$.

\subsubsection{$n$ qubits}
\label{nqubits}

We have hopefully clarified what is meant by a qubit. We will now
expand on our knowledge of quantum information by explaining what
people mean by having or transmitting $n$ qubits. We already know
that $n$ qubits is nothing more than a fancy way of saying $n$ two
level quantum systems. So the point is really to understand the
features displayed by the joint system of $n$ two level quantum
systems, possibly interacting with one-another. In section
\ref{BipartiteEntanglement}, we saw that, when you abandon the
safe territory of single particle quantum mechanics, you
immediately stumble over the remarkable phenomena of quantum
entanglement that make the quantum description of a composite
object very different from its classical description. Please note
that we are not contrasting macroscopic objects obeying the laws
of classical physics (say three beams of light), with microscopic
objects obeying the laws of quantum mechanics (say three photons).
Instead, we are remarking that even if you choose macroscopic
objects, say three beams of light, and you decide never to mention
the word photon, you will still be able to come out with states of
the joint macroscopic system that are entangled and therefore
completely beyond classical intuition. Let us be even more
explicit. Imagine that you have a classical physicist right in
front of you and you ask him the following question:

{\bf You}: How many complex numbers do you need to provide in
order to specify the joint state of a system comprised of three
polarized beams of light?

The classical physicist will probably find the expression {\it
joint state} rather peculiar, but he will still answer your
question on the basis of his knowledge of classical
electrodynamics.

{\bf Classical physicist}: To {\em completely} describe the state
of a composite system (ie. one composed of many subsystems) you
just need to specify the state of each subsystem individually. So
if you have $n$ arbitrary polarized light beams, you need $2 n$
complex numbers to describe {\em completely} the joint system, $2$
complex parameters for each of the $n$ systems. In fact the state
of each beam of light can be described by a superposition of say
horizontally and vertically polarized components.

\begin{equation}
\ket{\theta}= A_V e^{i\theta_V}\ket{V}+ A_H e^{i\theta_H}\ket{H}
\label{light}
\end{equation}
What we mean is only to prepare a beam of
light in a superposition of horizontally and vertically polarized
components. Instructions given in this form should be
understandable by a classical physicist, too. Furthermore the two
complex coefficients in equation \ref{light} can be interpreted as
follows: $A_V$ and $A_H$ are the moduli of the amplitude,
corresponding to the field strength, and $\theta_V$ and $\theta_H$
are the phases of the vertically and horizontally polarized
components. An example is light that is polarized at a $45$ degree
angle, which can also be viewed as an equally weighted
superposition of horizontally and vertically polarized light with
the same phase. The description of three such beams of light will
$obviously$ require $2 \times 3$ complex parameters.

\begin{figure}[!htb]
    \centerline{\epsfxsize=9.cm \epsffile{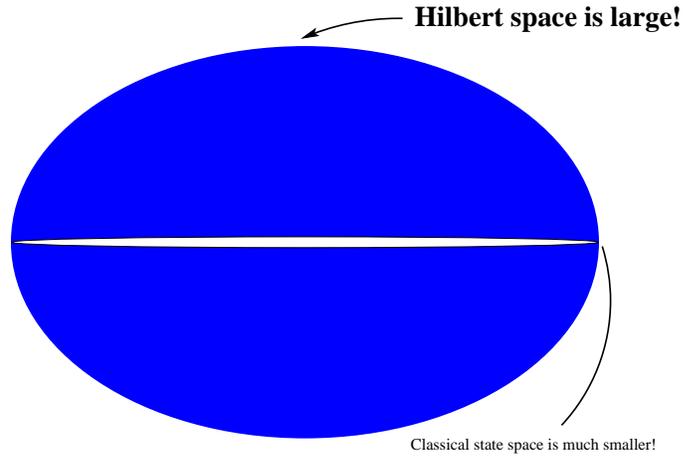}}
    \vspace*{0.15cm}
    \caption{Schematic picture of the whole Hilbert space, including entangled
    states, and the smaller space comprising only the disentangled
    states expected by a classical physicist.}
    \label{spaces}
\end{figure}

Unfortunately, statements that seem obvious sometimes turn out to
be wrong. The reader, who remembers our discussion of entanglement
in section \ref{BipartiteEntanglement}, may see where the problem
with the argument above lies. In order to describe an $n$-partite
object quantum mechanically, you need an enlarged Hilbert space
spanned by $2^{n}$ orthogonal state vectors. For example the joint
state of three beams of light is an arbitrary superposition of the
$2^{3}$ orthogonal state vectors, and therefore requires $8$
complex coefficients, not 6. Why $8$? Consider the state vector
$\ket{HHV}$ representing the state in which the first and the
second beams are horizontally polarized whereas the third is
vertically polarized. Here we used H and V, rather than 1 and 0 as
in section \ref{BipartiteEntanglement}, but the logic is the same.
How many of those vector states can you superpose? Well, each of
the three entries in $\ket{...}$ can be either H or V so you have
$2 \times 2 \times 2$ possibilities. Therefore any quantum state
can be written as the superposition of these 8 vectors in an 8
dimensional Hilbert space. However, as we saw in section
\ref{BipartiteEntanglement}, not every vector can be factorized in
three 2-dimensional vectors each describing a single beam of
light. If he insists on using only 6 parameters to describe a
tripartite system, the classical physicist will ignore many valid
physical states that are entangled! You may wonder how big that
loss is. In other words, how much of the Hilbert space of a
$n$-partite system, is actually composed of entangled states. The
answer is pretty straightforward. Product states predicted by
classical thinking "live" in a subspace of dimension $2 \times n$,
whereas the dimension of the whole Hilbert space for the joint
state of $n$ beams of light has $2^{n}$ dimension. Formally
stated, the phase space of a quantum many body system scales
exponentially with the number of components if you allow for
entanglement among its parts. The classical product states instead
occupy only an exponentially small fraction of its Hilbert space
as shown in figure \ref{spaces}.

Going back to our starting point, we say that we are able to hold
and manipulate $n$ qubits when we can prepare and keep $n$ beams
of light, $n$ two-level atoms or $n$ spin-$\frac{1}{2}$ particles
in a joint state $\ket{\psi}$ given by any arbitrary superposition
of the $2^{n}$ state vectors which can take the form

\begin{equation}
    |\psi\rangle = \sum_{i_1,\ldots,i_n=0}^{1} \alpha_{i_1\ldots i_n} |i_1\ldots
    i_n\rangle
\end{equation}
with $2^n$ complex amplitudes $\alpha_{i_1\ldots i_n}$. The actual
preparation of such a state presents a tremendous experimental
task no matter which constituents subsystems you choose. You need
to carefully control and "engineer" the interaction among all the
constituent components to choose the state you want and at the
same time you have to protect the joint state against
environmental noise. To date, this is possible with only a few
qubits and many people are skeptical about radical improvements in
the near future. The prospect of implementing quantum computation,
that requires manipulation of many qubits to be effective, seems
far beyond present capabilities.

\subsection{The quantum information content of a quantum system in qubits}
\label{Thequantuminformationcontentofaquantumsysteminqubits}

We want to make up for the pessimistic tone that ended the last
section with the discussion of an interesting feature of quantum
information that might be useful in case devices based on quantum
information theory are ever built. We will explain how an
arbitrary quantum state of a composite system comprised of $n$
interacting 2-level atoms, can be compressed and transmitted by
sending a number $m<n$ of qubits. As advertised in chapter 3, this
procedure justifies the use of the qubit as the unit of quantum
information and by analogy with classical data compression partly
justifies the otherwise misleading name $qubit$. We proceed in
close {\em mathematical} analogy to the classical case studied in
section \ref{Theinformationcontentofaclassicalstateinbits} and see
how well we can compress quantum states, ie. how many qubits are
needed to describe a quantum state. We first give a simple
example, that illustrates the key ideas, and then we reiterate
these ideas in a slightly more general and formal way.

\subsubsection{Quantum data compression: a simple example}
\label{Quantumdatacompressionasimpleexample}

Let us begin with the following very simple example, which is in
fact essentially classical, but displays all the relevant ideas of
the more general case. Consider a quantum source that emits
two-level systems with probability $p_0=0.95$ in state $|0\rangle$
and with probability $p_1=1-p_0=0.05$ in the orthogonal state
$|1\rangle$. Our knowledge of this preparation procedure for a
single qubit is represented by the density operator ${\hat \rho}$
given by

\begin{equation}
{\hat \rho}= 0.95 \projector{0} + 0.05 \projector{1} \label{ex}
\end{equation}

Note, that the two states generated by the oven have been chosen
to be orthogonal for simplicity. We will consider the more general
case later. For the time being, let us consider blocks of $7$
qubits generated by the source described above. Clearly any
sequence of qubits in states $|0\rangle$ and $|1\rangle$ is
possible, but some are more likely than others. In fact, typically
you will find either a sequence that contains only qubits in state
$|0\rangle$ or sequences with a single qubit in state $|1\rangle$
and all others in state $|0\rangle$, as shown below:

\begin{eqnarray}
    |\psi_{000}\rangle &=&
    |0\rangle|0\rangle|0\rangle|0\rangle|0\rangle|0\rangle|0\rangle \nonumber\\
    |\psi_{001}\rangle &=&
    |0\rangle|0\rangle|0\rangle|0\rangle|0\rangle|0\rangle|1\rangle \nonumber\\
    |\psi_{010}\rangle &=&
    |0\rangle|0\rangle|0\rangle|0\rangle|0\rangle|1\rangle|0\rangle \nonumber\\
    |\psi_{011}\rangle &=&
    |0\rangle|0\rangle|0\rangle|0\rangle|1\rangle|0\rangle|0\rangle \nonumber\\
    |\psi_{100}\rangle &=&
    |0\rangle|0\rangle|0\rangle|1\rangle|0\rangle|0\rangle|0\rangle \\
    |\psi_{101}\rangle &=&
    |0\rangle|0\rangle|1\rangle|0\rangle|0\rangle|0\rangle|0\rangle \nonumber\\
    |\psi_{110}\rangle &=&
    |0\rangle|1\rangle|0\rangle|0\rangle|0\rangle|0\rangle|0\rangle \nonumber\\
    |\psi_{111}\rangle &=&
    |1\rangle|0\rangle|0\rangle|0\rangle|0\rangle|0\rangle|0\rangle\;\; .\nonumber
\end{eqnarray}

The probability that you will get one of the above sequences is
$p_{likely}= (0.95)^{7}+ 7 (0.95)^{6} (0.05)= 0.955$. Of course,
these 'typical' states can be enumerated using just three binary
digits, i.e. $3$ binary digits are sufficient to enumerate
$95.5\%$ of all occurring sequences. This procedure is analogous
to labeling the typical sequences of 0s and 1s shown in figure
\ref{Shannon} except that we now 'enumerate' the typical sequences
with 'quantum states'. Now, let us see how we can use this fact
quantum mechanically. We define a unitary transformation that has
the following effect:

\begin{eqnarray}
    U|0\rangle|0\rangle|0\rangle|0\rangle|0\rangle|0\rangle|0\rangle &=&
    |0\rangle|0\rangle|0\rangle|0\rangle|0\rangle|0\rangle|0\rangle \nonumber\\
    U|0\rangle|0\rangle|0\rangle|0\rangle|0\rangle|0\rangle|1\rangle &=&
    |0\rangle|0\rangle|0\rangle|0\rangle|0\rangle|0\rangle|1\rangle \nonumber\\
    U|0\rangle|0\rangle|0\rangle|0\rangle|0\rangle|1\rangle|0\rangle &=&
    |0\rangle|0\rangle|0\rangle|0\rangle|0\rangle|1\rangle|0\rangle \nonumber\\
    U|0\rangle|0\rangle|0\rangle|0\rangle|1\rangle|0\rangle|0\rangle &=&
    |0\rangle|0\rangle|0\rangle|0\rangle|0\rangle|1\rangle|1\rangle \nonumber\\
    U|0\rangle|0\rangle|0\rangle|1\rangle|0\rangle|0\rangle|0\rangle &=&
    |0\rangle|0\rangle|0\rangle|0\rangle|1\rangle|0\rangle|0\rangle \label{sequence1} \\
    U|0\rangle|0\rangle|1\rangle|0\rangle|0\rangle|1\rangle|0\rangle &=&
    |0\rangle|0\rangle|0\rangle|0\rangle|1\rangle|0\rangle|1\rangle \nonumber\\
    U|0\rangle|1\rangle|0\rangle|0\rangle|0\rangle|1\rangle|0\rangle &=&
    |0\rangle|0\rangle|0\rangle|0\rangle|1\rangle|1\rangle|0\rangle \nonumber\\
    U|1\rangle|0\rangle|0\rangle|0\rangle|0\rangle|1\rangle|0\rangle &=&
    |0\rangle|0\rangle|0\rangle|0\rangle|1\rangle|1\rangle|1\rangle \;\; .\nonumber
\end{eqnarray}

In this case the unitary transformation is a matrix that maps a
set of 8 orthogonal column vectors on another set of 8 orthogonal
vectors in a complex vector space of dimension $2^{7}$. The effect
of this unitary transformation is to compress the information
about the typical sequences into the last three qubits, while the
first four qubits are always in the same pure state $|0\rangle$
and therefore do not carry any information. However, when $U$ acts
on other, less likely, sequences it will generate states that have
some of the first four qubits in state $|1\rangle$. Now comes the
crucial step, we throw away the first four qubits and obtain a
sequence of three qubits:

\begin{eqnarray}
    |0\rangle|0\rangle|0\rangle|0\rangle|0\rangle|0\rangle|0\rangle
    &\rightarrow& |0\rangle|0\rangle|0\rangle \nonumber\\
    |0\rangle|0\rangle|0\rangle|0\rangle|0\rangle|0\rangle|1\rangle
    &\rightarrow& |0\rangle|0\rangle|1\rangle \nonumber\\
    |0\rangle|0\rangle|0\rangle|0\rangle|0\rangle|1\rangle|0\rangle
    &\rightarrow& |0\rangle|1\rangle|0\rangle \nonumber\\
    |0\rangle|0\rangle|0\rangle|0\rangle|0\rangle|1\rangle|1\rangle
    &\rightarrow& |0\rangle|1\rangle|1\rangle \nonumber\\
    |0\rangle|0\rangle|0\rangle|0\rangle|1\rangle|0\rangle|0\rangle
    &\rightarrow& |1\rangle|0\rangle|0\rangle \nonumber\\
    |0\rangle|0\rangle|0\rangle|0\rangle|1\rangle|0\rangle|1\rangle
    &\rightarrow& |1\rangle|0\rangle|1\rangle \nonumber\\
    |0\rangle|0\rangle|0\rangle|0\rangle|1\rangle|1\rangle|0\rangle
    &\rightarrow& |1\rangle|1\rangle|0\rangle \nonumber\\
    |0\rangle|0\rangle|0\rangle|0\rangle|1\rangle|1\rangle|1\rangle
    &\rightarrow& |1\rangle|1\rangle|1\rangle
\label{sequence2}
\end{eqnarray}
Therefore we have compressed the $7$ qubits into $3$ qubits. Of
course we need to see whether this compression can be undone
again. This is indeed the case, when these three qubits are passed
on to some other person, this person then adds four qubits all in
the state $|0\rangle$ and then applies the inverse unitary
transformation $U^{-1}$ and obtains the states in equation
\ref{sequence1} back. This implies that this person will
reconstruct the correct quantum state in at least $95.5\%$ of the
cases and he has achieved this sending only $3$ qubits. As we
showed in the classical case (see equation \ref{Shannon entropy}),
in the limit of very long blocks composed of $n$ qubits, our
friend will be able to reconstruct almost all quantum states by
sending only $n H(0.95)= 0.2864 n$ qubits. Note that this
procedure also works when we have a superposition of states. For
example, the state

\begin{equation}
    |\psi\rangle = \alpha
    |0\rangle|0\rangle|0\rangle|0\rangle|0\rangle|0\rangle|0\rangle +
    \beta |0\rangle|0\rangle|0\rangle|0\rangle|0\rangle|0\rangle|1\rangle
\end{equation}

can be reconstructed perfectly if we just send the state of three
qubits given below:

\begin{equation}
    |\psi\rangle = \alpha
    |0\rangle|0\rangle|0\rangle +
    \beta |0\rangle|0\rangle|1\rangle
\end{equation}

Therefore not only the states in equation \ref{sequence1} are
reconstructed perfectly, but also all superpositions of these
states.

A very similar procedure would work also when we have a source
that emits quantum states $|\psi_i\rangle$ with probabilities
$p_i$, giving rise to an arbitrary density operator $\rho=\sum_i
p_i |\psi_i\rangle\langle\psi_i|$. Unlike the example in equation
\ref{ex}, the states $|\psi_i\rangle$ can be $non-orthogonal$
states of a two level system so the resulting density matrix is
$not$ in diagonal form. In this slightly more complicated case,
the first step consists in finding the eigenvectors and
eigenvalues of $\rho$. As the eigenvectors to different
eigenvalues are orthogonal, we are then in the situation of
equation \ref{ex}.  We can immediately see that the number of
qubits that need to be sent, to ensure that the probability with
which we can reconstruct the quantum state correctly is
arbitrarily close to unity, is given by $n$ times the Shannon
entropy of the eigenvalues of $\rho$ which is in turn equal to the
von Neumann entropy $S(\rho)$. Since we can reconstruct the
quantum state ${\rho}^{\otimes n}$ of a system composed of $n$
qubits by sending only $n S(\rho)$ qubits, we say that $n S(\rho)$
is the quantum information content of the composite system.

\subsubsection{Quantum data compression via Landauer's principle}

One may wonder whether the efficiency of quantum data compression
can be deduced from Landauers principle and indeed this is
possible. Given a source that generates $|\psi_i\rangle$ with
probabilities $p_i$, and gives rise to a density operator
$\rho=\sum_i p_i |\psi_i\rangle\langle\psi_i|$ we know from
section \ref{Erasingclassicalinformationfrom quantum states} that
the entropy of erasure per qubit is given by $S(\rho) kln 2$. Now
let us assume that we could compress the quantum information in
state ${\hat\rho}^{\otimes n}$ to $n(S(\hat\rho)-\epsilon)$ qubits
where $\epsilon\ge 0$. The state of each of these qubits will be
the maximally mixed state ${\hat
\omega}=\frac{1}{2}\projector{0}+\frac{1}{2}\projector{1}$ because
otherwise we could compress it even further. We can then calculate
the entropy of erasure of the $n(S(\rho)-\epsilon)$ qubits in
state $\hat\omega$ and find of course $n(S(\rho)-\epsilon) S({\hat
\omega})kln2 = n(S(\hat\rho)-\epsilon) H(\frac{1}{2})kln2 =
n(S(\hat\rho)-\epsilon) kln2$. Therefore the total entropy of
erasure would be given by the total number of qubits times the
entropy of erasure for the qubits $n(S(\hat\rho)-\epsilon)\times
kln2$ which is less than $n S(\hat\rho) kln2$. This however,
cannot be, because Landauer's principle dictates that the entropy
of erasure cannot be less than $S(\rho) kln2$ if the compressed
states should hold the same amount of information as the
uncompressed states. Therefore, we arrive at a contradiction which
demonstrates that the efficiency of quantum data compression is
limited by the Von Neumann entropy $S(\rho)$, as classical data
compression is limited by the Shannon entropy. This is the answer
to the first part of the second searching question in
\ref{introduction}. We still need to find out whether this
similarity between classical and quantum information extends also
to the act of copying information.

\subsection{Quantum information cannot be copied}
\label{Quantuminformationcannotbecopied}

In this section, we use Landauer's erasure principle to argue that
unlike classical bits qubits cannot be copied. This result is
often termed the no-cloning theorem. The basis of our arguments is
a reductio ad absurdum. We show that if Bob can clone an unknown
state sent to him by Alice, then he can violate Landauer's
principle. The logical steps of this argument are discussed below
in the context of an example.

\begin{enumerate}
\item  Alice starts by encoding letter $0$ and $1$, occurring with equal probabilities,
in the non-orthogonal states $\ket{\psi_0}$ and $\ket{\psi_1}$

\begin {equation} 0 \longmapsto \ket{\psi_0}=\ket{\uparrow}
\label{encoding1}
\end{equation}

\begin {equation} 1 \longmapsto \ket{\psi_1}=\frac{1}{\sqrt{2}}\ket{\uparrow}+\frac{1}{\sqrt{2}}\ket{\downarrow}.
\label{encoding2}
\end{equation}

We can find the upper bound to the information transmitted per
letter by using Landauer's principle. As discussed in section
\ref{Howmanybits}, the minimum entropy of erasure generated by
thermalisation of the carriers' states is given by $S(\rho)$ where
$\rho$ represent the incomplete knowledge that we have of the
state of each carrier:

\begin{equation} \rho=\frac{1}{2}\projector{\psi_1}+\frac{1}{2}\projector{\psi_0} .
\label{average state}
\end{equation}

After working out the matrix corresponding to ${\hat \rho}$ and
plugging it in the formula \ref{Von Neumann} for the Von Neumann
entropy, we find that the entropy of erasure and therefore the
information is equal to 0.6008 bits. This is less than 1 bit
because the two states are non-orthogonal and the von Neumann
entropy is less that the Shannon entropy of the probability
distribution with which the states are chosen, i.e.
$H(\frac{1}{2})=log2=${\it 1 bit}.

\item  Alice sends the message states to Bob who has the task to decipher her message.
Bob is also informed of how Alice encoded her letters (but of
course he does not know the message!) and uses this information in
his guess. No matter how clever Bob is, he will never recover more
information than what Alice encoded (i.e. more than 0.6008 bits).

\item  Now let us assume that Bob owns a machine that can clone an arbitrary unknown
quantum state and he uses it to clone an arbitrary number of times
each of the message-states Alice sends to him.

\item However, if Bob can clone the state of the message-system,
then, upon receiving any of the two states $|\psi_0\rangle$ or
$|\psi_1\rangle$ he can create a copy. Since the probability of
receiving each state is $\frac{1}{2}$, Bob will end up holding
either two copies of the first $|\psi_0\rangle|\psi_0\rangle$ or
two copies of the second state $|\psi_1\rangle|\psi_1\rangle$. We
can compute the density operator that describes this situation
following the rules described in section \ref{Crashcourse}:

\begin{equation}
    \rho_{two copies} =
    \frac{1}{2}|\psi_0\rangle|\psi_0\rangle\langle\psi_0|\langle\psi_0|
    +
    \frac{1}{2}|\psi_1\rangle|\psi_1\rangle\langle\psi_1|\langle\psi_1|
\end{equation}

The density operator $\rho_{two copies}$ is represented a
$4\times4$ matrix. After finding the eigenvalues of this matrix we
can calculate its Von Neumann entropy $S(\rho_{two copies})$. This
is a measure of the classical information that Bob has about the
letter received after cloning. We find:

\begin{equation}
    S(\rho_{two copies}) = 0.8113 > 0.6008 \; .
\end{equation}

Therefore the information content of the state has increased and
if we would push this further and create infinitely many copies,
then Bob would perfectly distinguish between the two
non-orthogonal states and he could extract one bit of information
per letter-state received. This, however, is not possible as we
cannot extract more info than Alice has originally encoded.
\end{enumerate}

The no-cloning theorem represents one of the most striking
differences between classical and quantum information. We
therefore conclude this section on quantum information by
completing our answer to the second question posed in the
introduction. Quantum information can be compressed in the sense
described in section
\ref{Thequantuminformationcontentofaquantumsysteminqubits}, but it
cannot be copied as we routinely do with classical information.

\section{Entanglement revisited}
\label{Entanglement}

In the last section, we have always encountered the concept of
entanglement as one of the central theme in quantum information
theory. However, we never systematically addressed the question of
what physical properties make entangled states peculiar and how
they can be engineered and exploited for practical purposes in the
lab. We now embark on this task. Our approach here will be based
on worked out examples. We have chosen the same approach and
numerical examples as in reference \cite{Mabuchi notes}, so that
the reader who masters the topics presented here can easily jump
to a more comprehensive and mathematical treatment. Throughout the
following sections, we concentrate exclusively on bipartite
entanglement for which a sufficient understanding has been
reached.

\subsection{The ebit}
\label{Theebit}

In section \ref{BipartiteEntanglement}, we saw that any arbitrary
superposition of the basis vectors ($\ket{01}$,
$\ket{11}$, $\ket{00}$,
$\ket{10}$) represents the physical state of a
bipartite system. So that must be true also for the vector
$\ket{\sigma_{AB}}$ given by:
\begin {equation}
    \ket{\sigma_{AB}}=\alpha\ket{01}+\beta\ket{10}.
    \label{entanglement1}
\end{equation}
where $\alpha$ and $\beta$ are two arbitrary $complex$ numbers
such that $|\alpha|^{2}+|\beta|^{2}=1$. We quickly remind the
reader that, according to the rules of quantum mechanics,
$|\alpha|^{2}$ is the probability for finding the first system
in $\ket{0}$ and the second in state $\ket{1}$ after a measurement, whereas
$|\beta|^{2}$ is the probability of finding the first system
in state $\ket{1}$ and the second $\ket{0}$. The states of systems $A$ and
$B$ are clearly anti-correlated. But this is not the whole story.

We remind the reader that what is remarkable about $\ket{\sigma_{AB}}$
is that it is
impossible to write it as a product state. The state
$\ket{\sigma_{AB}}$ is represented by a vector in the enlarged
Hilbert space $H_{AB}$ that cannot be factorised as the tensor
product of two vectors in $H_A$ and $H_B$. Therefore, we reach the
conclusion that $\ket{\sigma_{AB}}$ does represent the state of a
bipartite system, but we cannot assign a definite state to its
constituent components. In fact, even the terminology {\it
constituent components} is a bit misleading in this context. We
emphasize that the systems $A$ and $B$ can be arbitrary far from
each other but nevertheless constitute a single system. The
entanglement of the bipartite state $\ket{\sigma_{AB}}$ is then a
measure of the non-local correlations between the measurement
outcomes for system A and system B alone. These correlations are
the key to the famous Bell inequalities and origin of much
philosophical and physical debate \cite{Hardy 98} and more
recently the basis for new technological applications \cite{Ekert
J 96,Vedral P 98,Steane 98,Hughes ADLMS 95,PhysicsWorld
98,TMRQIT,Plenio V 98,Shor 99,PhysicsWorld 98,Bennett BCJPW
93,Bennett B 84,Hughes ADLMS 95}

\begin{figure}[!htb]
    \centerline{\epsfxsize=9.cm \epsffile{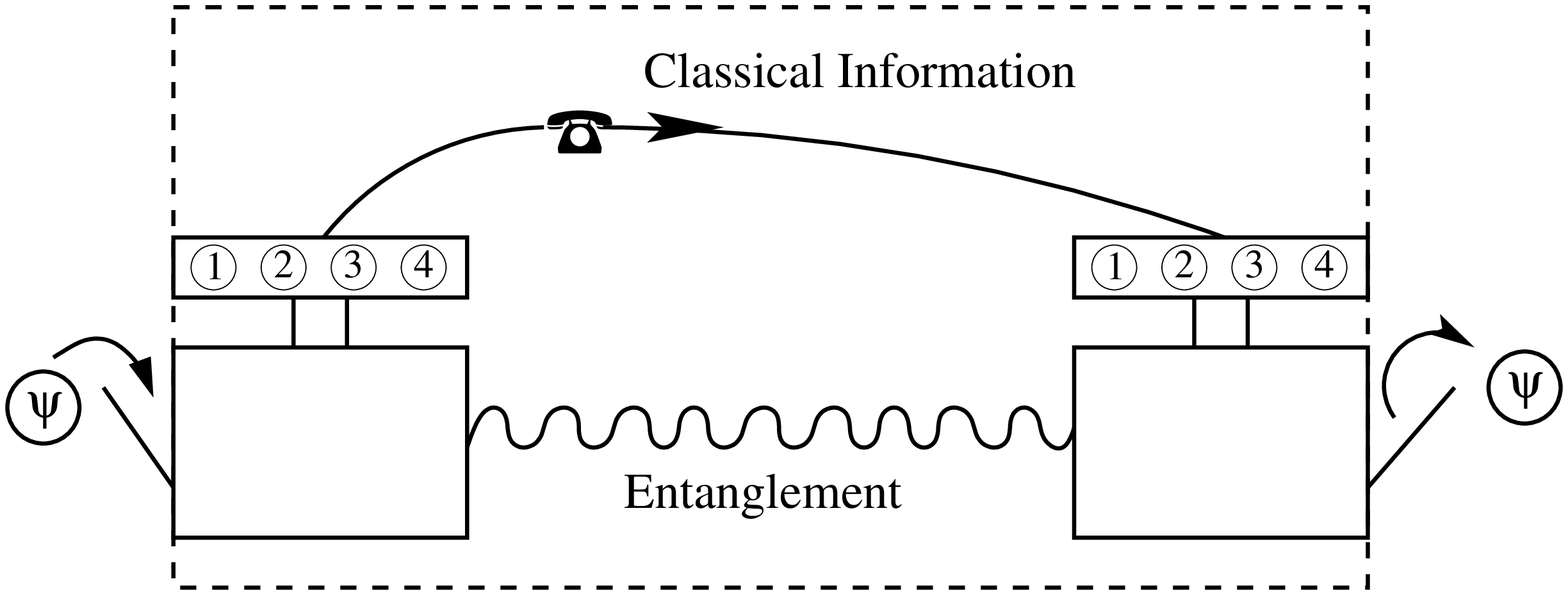}}
    \vspace*{0.15cm}
    \caption{A schematic picture of quantum state teleportation. A qubit
    in an unknown quantum state is entered into a machine which
    consumes one unit of entanglement (ebit) and a local measurement
    whose four possible outcomes are transmitted to the receiver. As a result
    the original state of the qubit is destroyed at the senders location and
    appears at the receivers end. The mathematical details can be found in
    \protect\cite{Bennett BCJPW 93,Plenio V 98}}
    \label{quantum fax}
\end{figure}

A basic question that arises in this context is how much
entanglement is contained in an arbitrary quantum state? A general
answer to this question has not been found yet, although quite a
lot of progress has been made \cite{Plenio V 98,Bennett BPS
96,Bennett DSW 96,Vedral P 98b} . In this article we confine
ourselves to the simplest case of bipartite entanglement for which
an extensive literature exists. As a first step we define the unit
of entanglement for a bipartite system as the amount of
entanglement contained in the maximally correlated state:
\begin{equation}
    \ket{\sigma_{AB}} = \frac{1}{\sqrt {2}}\ket{10}+
    \frac{1}{\sqrt{2}}\ket{01} \; .
    \label{ebit}
\end{equation}
We call this fundamental unit the ebit in analogy with the qubit
and the bit. Note that this state differs from the maximally
correlated state $\ket{\psi_{AB}}$ in equation \ref{maxent}, only
by a local unitary transformation and should therefore contain the
same amount of entanglement. The reason behind the name ebit will
be clear after reading section \ref{Entanglementpurification},
where we explain how to turn any multipartite entangled systems
into a group of $m$ ebits plus some completely disentangled
(product) states, just by using local operations and classical
communication. There is another reason, related to communication,
for choosing state Eq. (\ref{ebit}) as the unit of entanglement.
One can show that the ebit is the minimal amount of entanglement
that allows the non-local transfer of one unit of quantum
information. Such a procedure is quantum teleportation of one
qubit of quantum information \cite{Bennett BCJPW 93,Plenio V 98}.
For our purposes this process can be compared to the working of a
hypothetical quantum fax machine (see figure \ref{quantum fax}).
Alice, who is very far away from Bob can transmit the {\em unknown}
quantum state of a qubit to Bob by using this device. In what
follows, we regard the quantum fax machine as a black box (figure
\ref{quantum fax}). We are not interested in the internal
mechanism of this device nor in the procedures that Alice and Bob
have to learn to make it work. All we are interested are the
resources that this machine exploits and of course the result that
it produces. It turns out that the only two resources needed to
send the unknown quantum state of ONE qubit from Alice to Bob are:

\begin{enumerate}

\item ONE maximally entangled pair of particles shared between
Alice and Bob (represented by a wiggled line in figure
\ref{quantum fax}). For example, Bob is holding system B and Alice
system A and the joint state is $\ket{\sigma_{AB}}$ in equation
\ref {ebit}.

\item TWO classical bits that Alice must send to Bob through a
classical channel like an ordinary phone (represented by the
telephone line in figure \ref{quantum fax}).

\noindent
\end{enumerate}

\noindent If these two resources are available Alice and Bob can
successfully transmit the unknown quantum state of a qubit. The
existence of such quantum fax machines suggests that the sending
of 1 qubit can be accomplished by 1 ebit plus 2 classical bits.

There is an important difference between the quantum and classical
fax machine. After Alice sends the qubit to Bob the state of her
qubit (the original copy of the quantum message) gets destroyed.
Only one qubit survives the process and is in Bob's hands.
Incidentally also the ebit that acted as a sort of quantum channel
during the communication is destroyed. Those who were thinking of
buying a quantum fax machine and use it also as a quantum
photocopier will be disappointed. The reason for this is the no
cloning theorem \cite{Zurek W 81} discussed in section
\ref{Quantuminformationcannotbecopied}. Furthermore if we could
clone we would violate the law of the non-increase of entanglement
under local operations \cite{Plenio V 98} that we will explore in
the next few sections.

\subsection{Classical versus quantum correlations}

In the last section we mentioned that bipartite entanglement is a
measure of quantum correlations between two spatially separated
parts. We now want to make clear what is meant by quantum and
classical correlations in the context of an example.

Consider an apparatus that generates two beams of light in the
mixed state ${\hat \rho_{AB}}$ given by:

\begin{equation}
{\hat\rho_{AB}} =
\frac{1}{2}\projector{HH}+\frac{1}{2}\projector{VV}
\label{classical correlation}
\end{equation}
The notation above represents our incomplete knowledge of the
preparation procedure, namely the fact that we know that the two
beams were prepared either both vertically polarized or both
horizontally polarized but we do not know which of these two
alternatives occurred. If we perform a polarization measurement on
these two beams by placing the polarizer along the axis of
vertical or horizontal polarization we will find half of the time
the two beams both polarized in the vertical direction and half of
the time in the horizontal direction. In this sense, the
measurement outcomes for the two beams are $maximally$ correlated.
We say that mixed states like ${\hat \rho_{AB}}$ are classically
correlated. The adjective classical is there not because the
systems considered are necessarily classical macroscopic objects,
but rather because the origin of this correlation can be perfectly
explained in terms of classical reasoning. It simply arises from
our lack of complete knowledge of the preparation procedure.

If we represent the distinguishable single beam states $\ket{H}$
and $\ket{V}$ as the orthogonal column vectors $\left(\begin{array}{c}  1 \\
0 \end{array} \right)$ and $\left(\begin{array}{c}  0 \\ 1
\end{array} \right)$, respectively, we can then write the state ${\hat
\rho_{AB}}$ in matrix form following the guidance provided in
equations \ref{Tensor2}, \ref{projent} and \ref{mix1}

\begin{equation}
    {\hat \rho_{AB}}=
    \frac{1}{2}\left( \begin{array}{c c c c}
    1 & 0 & 0 & 0 \\
    0 & 0 & 0 & 0 \\
    0 & 0 & 0 & 0 \\
    0 & 0 & 0 & 1
\end{array} \right) \; .
\label{classical correlation 2}
\end{equation}

We now turn our attention to the maximally entangled state
$\ket{\psi_{AB}}=\frac{1}{\sqrt{2}}\ket{HH}+\frac{1}{\sqrt{2}}\ket{VV}$. When
the two beams are prepared in this pure state the outcomes of a polarization
measurement along the vertical and horizontal directions are maximally
correlated as in the previous case. However, there is an important difference
between the two. The maximally entangled state is a pure state. That means
there is nothing more that we can in principle know about it than what we can
deduce from its wave-function. So the origin of this correlation is not lack of
knowledge, because for a pure state we have complete information on the
preparation procedure. The state $\ket{\psi_{AB}}$ can be represented
mathematically using the same conventional choice of basis vectors and
following the same hints as the density matrix:

\begin{equation}
    \projector{\psi_{AB}}= \frac{1}{2}\left( \begin{array}{c c c c}
    1 & 0 & 0 & 1 \\
    0 & 0 & 0 & 0 \\
    0 & 0 & 0 & 0 \\
    1 & 0 & 0 & 1
\end{array} \right) \;\; .
\label{quantum correlation}
\end{equation}

A quick look at the entries of the matrix above shows that ${\hat
\rho_{AB}}$ is indeed a different mathematical object than
$\projector{\psi_{AB}}$. But this mathematical difference on paper
means nothing if we cannot interpret it physically. In other
words, how can you distinguish in the lab these two states from
each other, if they seem to have the same measurement statistics?
The answer is: turn the polarizer and measure again!
Unfortunately, we cannot perform this crucial experiment in front
of the reader but we can try to model it on paper and predict the
results on the basis of our knowledge of measurement theory as
developed in section \ref{Crashcourse}.

For example imagine that you turn the polarizer by $45^o$. Now you
have two new orthogonal directions that you can label $x$ and $y$.
These new directions are analogous to the directions of horizontal
and vertical polarization considered before.

The new polarization states can be expressed in terms of the
old ones by using simple vector decomposition:

\begin{equation}
    \ket{X}=\frac{1}{\sqrt{2}}\ket{V}+\frac{1}{\sqrt{2}}\ket{H}
    \label{X1}
\end{equation}

\begin{equation}
    \ket{Y}=\frac{1}{\sqrt{2}}\ket{V}-\frac{1}{\sqrt{2}}\ket{H}
    \label{X2}
\end{equation}

It seems natural to ask the question: are the measurement outcomes
of the two beams still maximally correlated (ie. the beams are
both found in either state $\ket{X}$ or state $\ket{Y}$? To answer
this question, we can check whether there is a non-vanishing
probability of finding one of the beams in state $\ket{X}$ and the
other in state $\ket{Y}$. To do that we have to first
construct the column vectors representing $\ket{X}$ and $\ket{Y}$
(see equation \ref{psi1}), then the single beam projectors
$\projector{X}$ and $\projector{Y}$ (see equation \ref{Mabuchi5.3}) and
finally the joint projector ${\hat P}$ given by $\projector{X}
\otimes \projector{Y}$ (see equation \ref{tensor projector}). We will not deprive
the reader from the pleasure of explicitly constructing the $4
\times 4$ matrix representing ${\hat P}$, a task well within
reach if one follows the hints given above. Once you have
${\hat P}$, you can calculate the probabilities of finding the
two beams anti-correlated in the new basis (ie. when you measure
with the polarizer turned by $\frac{\pi}{2}$) for both the
classically and quantum correlated states
($Prob^{\rho}_{anticorrelated}$ and
$Prob^{\psi}_{anticorrelated}$). Note that turning the polarizer
affects the measurement not the preparation procedure of states
${\hat \rho_{AB}}$ and $\projector{\psi_{AB}}$ that must be
prepared exactly as before. By using equation \ref{prob proj} we then find:

\begin{equation}
Prob^{\rho}_{anticorrelated}= tr\{{\hat P}{\hat
\rho_{AB}}\}=\frac{1}{4}.
\label{prob1}
\end{equation}

\begin{equation}
Prob^{\psi}_{anticorrelated}= tr\{{\hat
P}\projector{\psi_{AB}}\}=0.
\label{prob2}
\end{equation}

The results above demonstrate that the the two states in eq.
\ref{classical correlation} and \ref{quantum correlation} possess
different forms of correlations which we revealed by going from
the 'standard' basis to a rotated basis. This trick is the basis
for the formulation of Bell inequalities \cite{Hardy 98} which
show that a combination of correlations measured along different
rotated axes cannot overcome a certain value when the state on
which they are measured is classically correlated. If we measure
the same set of correlations on a quantum mechanically entangled
state, then this limit can be exceeded and this has been confirmed
in experiments.

\subsection{How to create an entangled state?}
\label{Howtocreate}

Another way to gain an intuitive understanding of the differences
between quantum and classical correlations is to investigate the
preparation procedures of states $\ket{\psi_{AB}}$ and ${\hat
\rho_{AB}}$. The latter can be generated by two distant parties,
Alice and Bob, who have a beam of light each and are allowed only
1) local operations on their own beam and  2) classical
communication via an ordinary phone. The entangled state instead
cannot be created unless Bob and Alice let their beams interact.
More explicitly, suppose that Alice and Bob are $both$ given each
one beam of light and are asked  to create first the mixed state
${\hat \rho_{AB}}$ and then the pure entangled state
$\ket{\psi_{AB}}$. What operations are they going to do, if they
start with the same resources in the two cases?

Let's first consider ${\hat \rho_{AB}}$. Alice who is in London
phones Bob who is in Boston and tells him to prepare his beam
horizontally polarized. That amounts to sending one bit of
classical information (ie. either H or V). Then she prepares her
beam also horizontally polarized. After completing this operation
the two have constructed the product state $\ket{HH}$. Now they
repeat the same procedure many time and each time they store their
beams in two rooms (one in London and the other in Boston) clearly
labeled with the $SAME$ number (for example, "experiment 1") and
with an H to indicate that the beam is horizontally polarized.
After doing this for $n$ times, they perform an analogous
procedure to create $\ket{VV}$ and they fill other $n$ rooms
carefully labeled with the same system, but they write V rather
than H, to indicate that they store vertically polarized beams.
Now, the two decide to erase the letter H or V from each room but
they keep the labeling number. After the erasure, Alice and Bob
have an incomplete knowledge of the state of the two beams
contained in each pair of room labeled with the same number. They
know that the two beams are either in state $\ket{HH}$ or
$\ket{VV}$ but they do not know which. The information the two
hold on each of the pair of correlated beams contained in rooms
labeled with the same number is correctly described by ${\hat
\rho_{AB}}$. They have in fact created an ensemble of pair of
beams in state ${\hat \rho_{AB}}$ by acting locally and just using
phone calls. The example above is a bit of a "theorist's
description of what is going on in the lab". The example captures
the crucial fact that classical correlations arise from 1) local
manipulations of the quantum states and 2) erasure of information
that in principle is available to some more $knowledgeable$
observers.

The situation is very different when Alice and Bob want to create
an entangled state and they start with two completely disentangled
product states like one beam in Boston and another independent one
in London. In this situation, one of the two has to take the plane
and bring his or her beam to interact with the other. Only at that
point can entanglement be created. In fact, one of the basic
results of quantum information theory is that {\it the net amount
of entanglement in a system cannot be increased by using classical
communication and local operations only}. So, if Alice and Bob
start with no entanglement at all, then they are forced to bring
the two beams together and let them interact in order to create
entanglement. We now would like to illustrate an example of two
beams that are initially in a disentangled state and become
entangled by interacting with each other. Suppose that Alice and
Bob hold a beam each polarized at an angle $\frac{\pi}{4}$ (see
equation \ref{X1}) . The two beams are initially far away from
each other so they are not interacting. The joint system can be
described mathematically by the product state $\ket{\psi_{AB}(0)}$
given below:

\begin{eqnarray}
    \ket{\psi_{AB}(0)}&=& (\frac{1}{\sqrt{2}}\ket{H}_A + \frac{1}{\sqrt{2}} \ket{V}_A)\otimes
    (\frac{1}{\sqrt{2}}\ket{H}_B + \frac{1}{\sqrt{2}} \ket{V}_B) \;\;.
    \nonumber\\
    &=&
        \frac{1}{2}\left( \begin{array}{c} 1 \\ 1 \end{array} \right)\otimes
        \left( \begin{array}{c}
         1 \\ 1
    \end{array} \right) \;\;.
    \nonumber\\
    &=&
    \frac{1}{2}\left( \begin{array}{c} 1  \\ 1  \\ 1 \\ 1 \end{array} \right) \;\; .
\label{initial}
\end{eqnarray}

The two beams in the product state $\ket{\psi_{AB}(0)}$ are
brought together and they start interacting with each other. The
time evolution of the original state is determined by the joint
Hamiltonian of the system ${\hat H}$ that is represented
mathematically by a $4 \times 4$ hermitian matrix because it has
to operate on vectors in the enlarged Hilbert space. Let us pick
up an Hamiltonian of this type, something easy so the calculation
does not get too complicated and let us see what happens.

\begin{equation}
    {\hat H}= \left( \begin{array}{c c c c}
    1 & 0 & 0 & 0 \\
    0 & 1 & 0 & 0 \\
    0 & 0 & 1 & 0 \\
    0 & 0 & 0 & -1
\end{array} \right) \;\; .
\label{hamiltonian}
\end{equation}

The basis vectors used to write the Hamiltonian ${\hat H}$ are the
same used to write $\ket{\psi_{AB}(0)}$ in equation \ref{initial}.
Since the matrix in equation \ref{hamiltonian} is diagonal, we can
read out the eigenstates and eigenvalues of the Hamiltonian. They
are the state vectors $\ket{HH}$, $\ket{HV}$, $\ket{VH}$ and
$\ket{VV}$ and the corresponding eigenvalues are equal to $1$,
$1$, $1$ and $-1$.

We can now write down the time evolution of the state
$\ket{\psi_{AB}(0)}$ by solving the Schr{\"o}dinger equation with
the Hamiltonian ${\hat H}$:

\begin{equation}
i\hbar\frac{\partial \psi_{AB}(t)}{\partial t}={\hat
H}\psi_{AB}(t) \label{SE}
\end{equation}

The Schr{\"o}dinger equation above is really a set of four linear
differential equations one for each component of the four
dimensional vector representing $\psi_{AB}(t)$. Usually, these
four differential equations would be coupled by the Hamiltonian so
you would have to diagonalize the corresponding matrix. In this
case however the Hamiltonian is already diagonal so we can redily
write the solution of this set of equations in vector form as:

\begin{equation}
\psi_{AB}(t)=\exp (\frac{-i}{\hbar}{\hat H}t)\psi_{AB}(0).
\label{Sol}
\end{equation}

The exponential of the Hamiltonian $\exp (\frac{-i}{\hbar}{\hat
H}t)$ is the diagonal matrix whose eigenvalues are the exponential
of the eigenvalues of the Hamiltonian's matrix (see equation
\ref{Von Neumann} and discussion below). The reader can also check
that this time evolution matrix is unitary. After time
$t=\frac{\pi \hbar}{2}$ (never mind the units) the matrix can be
written as:

\begin{eqnarray}
   \exp
(\frac{-i}{\hbar}{\hat H}t) &=& \left(
\begin{array}{c c c c}
    e^{\frac{-i \pi}{2}} & 0 & 0 & 0 \\
    0 & e^{\frac{-i \pi}{2}} & 0 & 0 \\
    0 & 0 & e^{\frac{-i \pi}{2}} & 0 \\
    0 & 0 & 0 & e^{\frac{+i \pi}{2}}
\end{array} \right) \;\; .
\nonumber\\
&=& \left( \begin{array}{c c c c}
    -i & 0 & 0 & 0 \\
    0 & -i & 0 & 0 \\
    0 & 0 & -i & 0 \\
    0 & 0 & 0 & i
\end{array} \right) \;\; .
\label{evolution}
\end{eqnarray}

According to equation \ref{Sol}, you can now write down the vector
$\psi_{AB}(t)$ just by multiplying the unitary matrix in equation
\ref{evolution} times the column vector $\psi_{AB}(0)$ given in
equation \ref{initial}. The result is that after time $\frac{\pi
\hbar}{2}$ the state vector representing the system is:

\begin{equation}
    \ket{\psi_{AB}(t)}= \frac{-i}{2}\left( \begin{array}{c} 1  \\ 1  \\ 1 \\ -1 \end{array} \right).
\label{final}
\end{equation}

You can check by inspection that the state in equation \ref{final}
is entangled (ie. it cannot be factorized). The more ambitious
reader may consult reference \cite{Mabuchi notes} that explains in
simple terms the systematic criteria to check whether the state of
a bipartite system is entangled or not in the context of this
example.

Whatever way you choose to convince yourself that the state above
is entangled the conclusion is the same. States that can be
factorized arise mathematically only for very special choices of
the entries of the corresponding vectors. Under Hamiltonian
evolution the value of these entries will change and in general it
will not be possible to factorize the state any more. The
discussion above shows that the process by which two independent
systems in a product state like $\ket{\psi_{AB}(0)}$ get entangled
is indeed quite natural provided that the two systems are brought
together and left to interact with each other. However, most
interaction will not lead to a maximally entangled state. It is
therefore important for applications like teleportation to devise
techniques by which one can distill a set of ebits from an
ensemble of partially entangled states like $\ket{\psi_{AB}(t)}$
in equation \ref{final}. This is the subject of the next section.

\subsection{Entanglement distillation}
\label{Entanglementpurification}

We emphasize that the fundamental law of quantum information
processing does not rule out the possibility to $occasionally$
increase the net amount of entanglement in a system by using local
operations and classical communication only, provided that $on$
$average$ the net amount of entanglement is not increased. This
implies that it should be possible to devise strategies to turn a
partially entangled pair of particles into an ebit provided that
this strategy sometimes leads to an increase and other times to a
loss of entanglement so that on average the "entanglement balance"
stays the same. We first consider a simple example of entanglement
distillation and then we look at the efficiency of a general
distillation procedure by using Landauer's principle.

\subsubsection{A simple example}
\label{entex}

Alice is still in London and Bob in Boston. They share a non
maximally entangled pair of particles in the state
$\ket{\psi_{AB}}= \alpha \ket{00} + \beta \ket{11}$, where
$\alpha$ $\neq$ $\beta$. They want to turn it into an ebit but
they are only allowed to act locally on their own particle but not
to let the two interact. Furthermore, their communication must be
limited to classical bits sent over an ordinary channel, nothing
fancy like sending or teleporting quantum states is allowed. The
reason why we demand such tough conditions on Bob and Alice and we
insist on them not to freely meet up is because we want to
investigate the issue of locality versus non-locality. This is
really the main theme behind our study of entanglement, so we have
to be extra careful in keeping track of what they do. That still
leaves a lot of room for manipulation on both Alice's and Bob's
side. For example the two can add other particles on their own
side and let them interact with the entangled particle they are
holding and perform measurements on them. We now describe what
operations the two perform in order to distill one ebit.

\begin{enumerate}

\item Alice adds another particle in state $\ket{0_A}$ on her side. Note
that the subscript $A$ denotes particles on Alice's side and $B$
on Bob's side. Now the joint state of the entangled pair plus the
extra particle is given by the product state $\ket{\psi_{tot}}$
given below:

\begin{equation}
\ket{\psi_{tot}} = \ket{0_A}\otimes(\alpha \ket{0_A}\ket{0_B} +
\beta \ket{1_A}\ket{1_B}) . \label{add}
\end{equation}

We can collect the states of the two particles on Alice side in
the same four dimensional column vector and rewrite equation
\ref{add} as:

\begin{equation}
\ket{\psi_{tot}} = \alpha \ket{00}_A\ket{0_B} + \beta
\ket{01}_A\ket{1_B} \label{add2}
\end{equation}

\item Now Alice performs a unitary transformation ${\hat U}$ on her two
particles. As we mentioned in the previous section, a unitary
transformation can be implemented by letting the joint system
evolve for a certain time as dictated by a suitably chosen
Hamiltonian (see example in equation \ref{evolution}). The unitary
transformation ${\hat U}$ that Alice needs to implement on the
joint state of her two particles is given below in matrix form:

\begin{equation}
    {\hat U} = \left( \begin{array}{c c c c}
    \frac{\beta}{\alpha} & 0 & -\frac{\sqrt{\alpha ^{2}-\beta ^{2}}}{\alpha} & 0 \\
    0 & 1 & 0 & 0 \\
    \frac{\sqrt{\alpha ^{2}-\beta ^{2}}}{\alpha} & 0 & \frac{\beta}{\alpha} & 0 \\
    0 & 0 & 0 & 1
\end{array} \right) \;\; .
\label{unitary2}
\end{equation}

The reader can check that, when the unitary transformation is
applied on her states $\ket{00}_A$ and $\ket{01}_A$, Alice
achieves the following:

\begin{eqnarray}
{\hat U}\ket{00}_A &=& \frac{\beta}{\alpha}\ket{00}_A+
\frac{\sqrt{\alpha ^{2}-\beta ^{2}}}{\alpha}\ket{10}_A \;\; ;
\nonumber\\ {\hat U}\ket{01}_A &=& \ket{01}_A\;\; .
\label{transform}
\end{eqnarray}

Hence, when the unitary transformation ${\hat U}$ is applied to
the joint state of the three particles $\ket{\psi_{tot}}$, the
state of the particle on Bob's side is unaffected whereas the
state of the two on Alice's side is changed according to equation
\ref{transform}:

\begin{eqnarray}
    {\hat U}\ket{\psi_{tot}} &=& \beta \ket{00}_A \ket{0_B}
    \nonumber\\
    &&+ \sqrt{\alpha ^{2}-\beta ^{2}}\ket{10}_A \ket{0_B}+
\beta\ket{01}_A \ket{1_B}. \label{tran1}
\end{eqnarray}

We can split Alice's vector states in equation \ref{tran1} and
isolate the state of the entangled pair from the state of the
particle added on Alice's side by writing the latter first in the
equation below:

\begin{eqnarray}
    {\hat U}\ket{\psi_{tot}} = \sqrt{2}\beta \ket{0_A}
    \frac{\ket{0_A}\ket{0_B}+\ket{1_A}\ket{1_B}}{\sqrt{2}}
    \nonumber\\
    +\sqrt{\alpha ^{2}-\beta ^{2}}\ket{1_A}\ket{0_A}\ket{0_B}\;\; .
\label{tran2}
\end{eqnarray}

\item Now, Alice decides to perform a measurement on the extra
particle she is holding on her side. She chooses the observable
that has $\ket{0}$ and $\ket{1}$ as its eigenstates. There are two
possible scenarios:

a) Alice finds the extra particle in state $\ket{0}$. Then the
total state is $\ket{0}_A \otimes \frac{1}{\sqrt{2}}(\ket{0_A
0_B}+\ket{1_A 1_B})$. Alice and Bob share a maximally entangled
state. This event occurs with probability $2\beta^{2}$.

b) Alice finds the extra particle in state $\ket{1}$. Then the
total state is $\ket{1_A} \otimes \ket{0_A 0_B}$. The procedure
was unsuccessful and the two lost their initial entanglement. This
possibility occurs with probability $1 - 2\beta^{2}$.

\item Alice phones Bob and informs him of the measurement outcomes.
If the procedure is successful Bob holds his particle otherwise
they try again.

\noindent
\end{enumerate}

\noindent A question that arises naturally in this context is the
following: what is the maximum number of ebits that Alice and Bob
can extract from a large ensemble of $N$ non maximally entangled
states? We will answer this question by using Landauer's erasure
principle.

\subsubsection{Efficiency of entanglement distillation from Landauer's
principle}
\label{entpurLand}

We start by considering an example of a process that will cause
two systems to become entangled: a quantum measurement. A quantum
measurement is a process by which the apparatus and the system
interact with each other so that correlations are created between
the states of the two. These correlations are a measure of the
information that an observer acquires on the state of the system
if he knows the state of the apparatus.

Consider an ensemble of systems $S$ on which we want to perform
measurements using apparatus $A$. A general way to write the
state of $S$ is
\begin{equation}
\left| \psi _{S}\right\rangle
=\frac{1}{\sqrt{N}}\sum_{i=1}^{N}\left| s_{i}\right\rangle ,
\end{equation}

\noindent where $\left\{ \left| s_{i}\right\rangle \right\} $ is
an orthogonal basis. In our previous example, the orthogonal basis
was given by the vertically and horizontally polarized states.
When the apparatus is brought into contact with the system the
joint state of $S$ and $A$ is given by
\begin{equation}
\left| \psi _{S+A}\right\rangle
=\frac{1}{\sqrt{N}}\sum_{i=1}^{N}\left| s_{i}\right\rangle \left|
a_{i}\right\rangle .
\end{equation}

\noindent The result of the act of measurement is to create
correlations (ie. entanglement) between the apparatus and the
system. The equation above is a generalization of equation
\ref{ebit}.

An observation is said to be imperfect when it is unable to
distinguish between two different outcomes of a measurement. Let
$A$ be an imperfect measuring apparatus so that $\left\{ \left|
a_{i}\right\rangle \right\} $ is NOT an orthogonal set. A
consequence of the non-orthogonality of the states $\ket{a_{i}}$
is that we are unable to distinguish with certainty the correlated
states $\ket{s_{i}}$. There is no maximal correlation between the
state of the system and the apparatus, which means that $S$ and
$A$ are not maximally entangled). However, suppose that by acting
locally on the apparatus we can transform the whole state $\left|
\psi _{S+A}\right\rangle$ into the maximally entangled state
$\ket{ \phi _{S+A}}$:

\begin{equation}
\left| \phi _{S+A}\right\rangle
=\frac{1}{\sqrt{N}}\sum_{i=1}^{N}\left| s_{i}\right\rangle \left|
b_{i}\right\rangle ,
\end{equation}

\noindent where $\left\{ \left| b_{i}\right\rangle \right\} $ IS
an orthogonal set. This does not increase the information between
the apparatus and the system since we are not interacting with the
system at all. In order to assess the efficiency of this
distillation procedure we need to find the probability with which
we can distill successfully.

The state of the apparatus only, after the correlations are
created, is given by the reduced density operator first
encountered in section \ref{Thereduced}:
\begin{equation}
tr_{S}(\left| \psi _{S+A}\right\rangle \left\langle \psi
_{S+A}\right| )=\rho _{A} .
\end{equation}

\noindent  Landauer's principle states that to erase the
information contained in the apparatus we need to generate in the
environment an entropy of erasure larger than $S(\rho _{A})$ and
this has to be greater than or equal to the information gain.
After we purify the state to $\left|
\phi _{S+A}\right\rangle $ with a probability $p$ , we gain $%
p\log N$ bits of information about the system. In fact, since we
have maximal correlations now, the result of a measurement enables
us to distinguish between N equally likely outputs. The rest of
the state contains no information because it is completely
disentangled and therefore there are no correlations between the
states of the system and the apparatus. After reading the state of
the apparatus we will not gain any useful knowledge on the state
of the system.

By Landauer's principle, the entropy of erasure is greater than or
equal to the information gain before purification and this is in
turn greater than or equal to the information the observer has
after purification, because the apparatus is not interacting with
the system so the information can not increase. We thus write

\begin{equation}
S(\rho _{A})\geq p\log N .
\end{equation}

\noindent The upper bound to purification efficiency is therefore
\begin{equation}
p\leq S(\rho _{A})/\log N .
\end{equation}

This bound obtained from Landauer's principle is actually
achievable as has been proven in \cite{Bennett BPS 96} by
construction of an explicit procedure that achieves it. It is
nevertheless satisfying that Landauer's principle is able to give
a sharp upper bound with a minimal amount of technicalities and by
doing so it provides an informal argument for using the Von
Neumann entropy as a measure of bipartite entanglement. With this
result we answer the last of the three questions posed in the
introduction that have served as guidelines for our exploration of
the physical theory of information.

\noindent

\section{Conclusion}

This is really the end of our long investigation on the properties
of entanglement, classical and quantum information. We hope to
have reasonably delivered what we promised in the introduction.
Throughout the paper, we used the pedagogical technique of going
backwards and forward among different aspects of the subject, each
time increasing the level of sophistication of the ideas and
mathematical tools employed. This method has the advantage of
allowing enough time for "different layers of knowledge to
sediment in the mind of the reader". Unfortunately, there is also
the inevitable side effect that a proper understanding of the
subject matter will only follow when the reader goes through the
material more than once. For example, the understanding of the
differences between quantum and classical information crucially
relies on the appreciation of the concepts of classical and
quantum correlations that were explicitly studied only at the end
of the article. No matter how hard we tried to argue with words
previously, a proper grasp of these topics came only after
employing more advanced mathematical tools developed in later
parts of the paper.

To prevent the reader from feeling lost, we will now attempt to
recap the content of the paper. In the first part, the scene was
dominated by the Shannon entropy that helped us to define and
evaluate the amount of classical information encoded in a
classical object or message. We were also able to find a bound on
the {\it classical information capacity of a noisy classical
channel} by using Landauer's principle. The answer depended once
again on the Shannon entropy. Following a brief recap of quantum
mechanics, our interest slightly shifted to quantifying the amount
of classical information encoded in quantum systems. This was
achieved by introducing the Von Neumann entropy. After developing
a suitable thermalization procedures to erase information from
quantum systems, we managed to employ Landauer's principle to
justify the Holevo bound. This bound expresses the {\it classical
information capacity of a noisy quantum channel} in terms of the
Von Neumann entropy. That completed our investigation of classical
information.

We then turned our attention to quantifying the amount of quantum
information encoded in a quantum object or message. This result,
which is based on quantum data compression, was obtained employing
Landauer's principle and provided a solid basis for the
introduction of the qubit as the fundamental unit of quantum
information. The answer to this question was once again given by
the von Neumann entropy. Quantum information can be compressed,
but unlike classical information, it cannot be copied. This was
our conclusion after studying the no-cloning theorem with the help
of Landauer's erasure principle.

Motivated by these successes we tried to shed light on the
phenomena of entanglement using Landauer's principle. We explained
that creating a pair of entangled states is not difficult after
all. Any two systems initially uncorrelated will get entangled
just by interacting with each other. However, it is not equally
easy to create quantum states that are $maximally$ entangled over
large distance. This problem can be overcome by designing suitable
distillation procedures by which maximally entangled states,
ebits, are produced from an ensemble of non-maximally entangled
states without increasing the total amount of entanglement. To
some extent this procedure provides a way to measure the amount of
entanglement (in ebits) contained in a system composed of {\em
only} two parts. The efficiency of a distillation procedure was
once again expressed in terms of the von Neumann entropy after
carrying out a simple analysis based on Landauer's principle. The
von Neumann entropy in quantum information theory is so widespread
to justify the claim that the whole field is really about its use
and interpretation , as classical information theory was based on
the Shannon entropy. \cite{Preskill notes}.

After reading this summary you might have noticed two glaring
omissions in our treatment. Firstly, we spent a lot of time
discussing the classical information capacity of a noisy classical
and quantum channel, but we never mentioned the more interesting
problem of the quantum information capacity of a noisy quantum
channel. In other words how many qubits can you send through a
noisy channel when the letters of your message are encoded in
arbitrary quantum states? Secondly, we never mentioned how to
generalize our discussion of entanglement measure to the useful
and interesting case of entangled states composed of more than two
particles.

We reassure the reader that these omissions are not motivated by
our compelling desire to meet the deadline for submission of this
paper, but rather by the fact that nobody really knows the answer
to these fundamental and natural questions. We do not know whether
one can push Landauer's principle to investigate these problems.
Landauer' principle is somehow limited to the erasure of classical
information whereas the questions above are completely quantum.
However, Landauer' principle can be used to yield upper bounds to
entanglement distillation a completely non-classical procedure.
Therefore the hope that Landauer's principle can shed some light
on these unsolved problems may not remain unfulfilled.

Anyway, these final remarks prove the point that, although a
large amount of work has been published since Shannon, there is
still room for further research in the foundations of information
theory. It is also evident that this research belongs to
fundamental physics as much as it does to engineering. If you
found some of the ideas in this paper fascinating and you wish to
start working in the field, you may want to start by studying some
further introductory texts such as \cite{Preskill notes,Plenio
notes,Mabuchi notes,Peres 95,Nielsen C 00}. Perhaps someday, we
will find out the answer to the questions above from you.

\noindent {\bf Acknowledgments. } We would like to thank Miles Blencowe, John
Calsamiglia, Susana Huelga, William Irvine, Daniel Jonathan, Peter Knight,
Polykarpos Papadopoulos, Stephen Parker, Vlatko Vedral and Shashank Virmani for
discussion on the topics involved in this article. This work was supported by
EPSRC, the Leverhulme Trust, the EQUIP project of the European Union and the
European Science Foundation programme on quantum information theory. The final
stages of this work have been carried out at the Erwin Schr{\"o}dinger
Institute in Vienna.

\end{multicols}
\newpage

\begin{center}
    Biographies
\end{center}

Martin Plenio studied in G{\"o}ttingen (Germany) where he obtained both his
Diploma (1992) and his PhD (1994) in Theoretical Physics. After his PhD he
joined the Theoretical Quantum Optics group at Imperial College with a
Feodor-Lynen grant of the Alexander von Humboldt Foundation as a postdoc. Since
January 1998 he is a Lecturer in the Optics Section of Imperial College. His
main research interests lie in Quantum Information Theory and Quantum
Optics.\\[2.cm]

Vincenzo Vitelli received his first degree in theoretical physics
(2000) from Imperial College. He is now a first year graduate
student at Harvard University. His main research interests lie in
the area of quantum information theory and statistical physics.

\end{document}